\documentclass[letterpaper,twocolumn,10pt]{article}
\usepackage{usenix2019_v3}

\usepackage{tikz}
\usepackage{amsmath}
\usepackage{CJKutf8} 
\usepackage{csquotes}
\usepackage{xspace}
\usepackage{multicol} 
\usepackage{multirow} 
\usepackage[export]{adjustbox} 
\usepackage{array,threeparttable,siunitx} 
\usepackage{tabularx} 
\usepackage{booktabs} 
\usepackage{lipsum}
\usepackage{colortbl}
\usepackage{float}
\newcommand{\st}{\textsuperscript{st}\xspace}
\newcommand{\nd}{\textsuperscript{nd}\xspace}
\newcommand{\rd}{\textsuperscript{rd}\xspace}
\renewcommand{\th}{\textsuperscript{th}\xspace}

\newcolumntype{C}[1]{>{\centering}m{#1}}

\newcommand{\mcthree}[1]{\multicolumn{3}{c}{#1}}

\newcommand{\Ca}{\cellcolor[RGB]{253,223,192}}
\newcommand{\Cb}{\cellcolor[RGB]{253,185,125}}
\newcommand{\Cc}{\cellcolor[RGB]{252,140, 59}}
\newcommand{\Cd}{\cellcolor[RGB]{233, 93, 13}}

\newcommand{\ts}{\textsuperscript}

\usepackage{enumitem}
\setlist{nosep}

\usepackage{longtable}

\usepackage{caption}

\usepackage{float}
\newfloat{copyright}{b} 
 
 \def\thecopyright{
 	\begin{copyright}[b] 
	\begin{center}
	\setlength{\unitlength}{1pc}
	\begin{picture}(20,3) 
	\put(0,-0.95){ 
	\parbox[b]{50pc}{\baselineskip 9pt 
	\includegraphics[width=3cm]{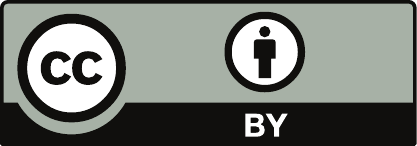}\par
	\footnotesize{Copyright is held by the authors.\\
	This is an extended version of our USENIX Security '23 paper~\cite{sombatruang_internet_2023}.}
	}}
	\end{picture}
	\end{center}
	\end{copyright}
}

\begin{document}

\date{}

\title{\Large \bf Internet Service Providers' and Individuals'\\ Attitudes, Barriers, and Incentives to Secure IoT}

\author{
{\rm Nissy Sombatruang\textsuperscript{1}}
\and
{\rm Tristan Caulfield\textsuperscript{2}}
\and
{\rm Ingolf Becker\textsuperscript{2}}
\and
{\rm Akira Fujita\textsuperscript{1}}
\and
{\rm Takahiro Kasama\textsuperscript{1}}
\and
{\rm Koji Nakao\textsuperscript{1}}
\and
{\rm Daisuke Inoue\textsuperscript{1}}
\and
\begin{minipage}{\textwidth}
\centering
\textsuperscript{1}National Institute of Information and Communications Technology\\
\textsuperscript{2}University College London\\
\end{minipage}
} 

\maketitle
\thecopyright{}

\begin{abstract}

Internet Service Providers (ISPs) and individual users of Internet of Things (IoT) play a vital role in securing IoT. However, encouraging them to do so is hard. Our study investigates ISPs' and individuals' attitudes towards the security of IoT, the obstacles they face, and their incentives to keep IoT secure, drawing evidence from Japan.

Due to the complex interactions of the stakeholders, we follow an iterative methodology where we present issues and potential solutions to our stakeholders in turn. 
For ISPs, we survey 27 ISPs in Japan, followed by a workshop with representatives from government and 5 ISPs. Based on the findings from this, we conduct semi-structured interviews with 20 participants followed by a more quantitative survey with 328 participants. We review these results in a second workshop with representatives from government and 7 ISPs. The appreciation of challenges by each party has lead to findings that are supported by all stakeholders.

Securing IoT devices is neither users' nor ISPs' priority. Individuals are keen on more interventions both from the government as part of regulation and from ISPs in terms of filtering malicious traffic. Participants are willing to pay for enhanced monitoring and filtering. While ISPs do want to help users, there appears to be a lack of effective technology to aid them. ISPs would like to see more public recognition for their efforts, but internally they struggle with executive buy-in and effective means to communicate with their customers. The majority of barriers and incentives are external to ISPs and individuals, demonstrating the complexity of keeping IoT secure and emphasizing the need for relevant stakeholders in the IoT ecosystem to work in tandem.

\end{abstract}
\section{Introduction} \label{Introduction}
The Internet of Things (IoT) brings several benefits to individuals and society. The ability to monitor properties remotely via smart CCTV, adjust heating remotely, and utilise energy more efficiently with a smart meter are just a few examples that are gradually becoming part of many people's lives. 
However, the lagging of security and privacy on these IoT devices can expose users and the supporting networks to cybersecurity risks. 
The 2016 Dyn attack which caused large-scale disruption to Internet services is a prime example of the risks that the IoT can pose to both individual users of IoT and ISPs~\cite{antonakakis2017understanding}.
It provides compelling evidence for the need for both parties to keep IoT secure.

The increasing involvement of ISPs in cleaning up infected home computers \cite{asghari2015economics,edwards2011role,cetin2018let,active,ccc} demonstrates that ISPs are key to securing IoT. 
Individuals, as the owner and the user of IoT devices, also have an important role in securing these devices. 
However, the attitudes of ISPs and individuals towards the need to secure IoT, and the incentives and obstacles that encourage or discourage them to act to secure IoT are still poorly understood.

This study seeks to understand these attitudes, obstacles, and incentives by investigating ISPs and individuals in Japan. Japan is of particular interest due to the country's fast-growing IoT adoption~\cite{japaniotstat, japaneu} and the government's ambition to keep the IoT ecosystem secure. This is most notably evident in the National Operation Towards IoT Clean Environment (NOTICE), an ongoing nationwide project to identify and remediate vulnerable and infected IoT devices~\cite{notice}. Our findings will improve the effectiveness of NOTICE and similar projects elsewhere.

For ISPs, we survey 27 ISPs across Japan followed by two workshops in which 5 and 7 ISPs participated. 
For individuals, we conduct semi-structured interviews with 20 participants and followed this up with an online survey with 328 participants.

Our findings shed light on the profound complexity of the effort to encourage ISPs and individuals to keep IoT secure.
While individuals have some concerns about the security and privacy of IoT devices and ISPs are concerned about their networks hosting infected/vulnerable IoT devices, keeping IoT secure is only a secondary priority for them.
Individuals and ISPs also faced various barriers that deterred them from doing so.

Most of the key incentives are external to ISPs and individuals; hence, the onus of implementing these incentives falls onto other stakeholders in the IoT ecosystem.
In the grand scheme of things, these findings suggest that: 1) solutions can not be unilateral on the part of one stakeholder, 2) good solutions require ISPs, governments, device manufacturers, and individuals to work together, and 3) the other stakeholders must help support and motivate individuals in their role. 

In summary, our contributions are as follows:
\begin{itemize}
\item We investigate ISPs' and individuals' attitudes, barriers, and incentives to secure IoT, being the first to undertake an integrated approach --- by examining three aspects from two stakeholders in one study.
\item We provide evidence of these attitudes, barriers, and incentives, the latter of which include the misaligned incentive between ISPs and individuals.
\item We synthesise lessons learned and propose considerations to encourage ISPs and individuals to secure IoT.
\end{itemize}

\section{Background and related work} 
\label{Background}

Along with its many benefits, the IoT also has disadvantages---one of the most criticised being the risks to security and privacy. Plenty of previous studies have produced evidence of the vulnerabilities found on IoT devices. For example, Liu et al~\cite{liu2019uncovering} uncovered security issues in one smart home system which allowed an attacker to compromise a passphrase guarding the communication over the local wireless network. 
Morgner et al~\cite{morgner2017insecure} showed how an attacker can exploit a vulnerability in Zigbee 3.0, a wireless technology used in devices such as door locks, and take over the devices from distance. 
Alrawi et al~\cite{alrawi2019sok} also evaluated and found vulnerabilities in a long list of IoT devices, concerning as some of these devices are popular products in the market. 

More frightening than the theoretical attacks is the mounting evidence of \textit{real-life} attacks on IoT devices.
Media reports about the hacking of IoT such as Internet-connected CCTV (e.g. \cite{cctvhackjapan, cctvhackuk, cctvhackus}), and smart home systems~\cite{smarthomehack} are not new today.
The most infamous IoT attack to date is the 2016 DYN attacks in which millions of IoT devices infected by Mirai, an IoT malware, were compromised and used to launch a distributed denial-of services, causing a large-scale disruption to Internet services~\cite{kaperskyiotdyn, antonakakis2017understanding}.

With the IoT market expected to grow exponentially---consumer spending is estimated to be 1.6 trillion US dollars by 2025~\cite{iotstat}---the need to secure IoT devices cannot be ignored.

\subsection{The role of ISPs in keeping IoT secure}

To understand ISPs' roles in keeping IoT secure, understanding their roles in mitigating botnets, a network of computers infected by malware, paves the foundation. 

In one of the earliest works in this area, Van Eeten et al~\cite{van2010role} analyzed a global set of spam data between 2005--2008 and showed that a small number of ISPs accounted for a significant percentage of unique IP addresses used for sending spam worldwide, demonstrating the ISPs' unique position as intermediaries in botnet mitigation.
Their subsequent work evaluating the role and performance of ISPs in botnet mitigation across 60 countries found that although the ISPs' performances varied, the ISPs can and do make a difference, especially in identifying, notifying, and quarantining the infected customer~\cite{asghari2015economics}. 

Pijpker and Vranken~\cite{pijpker2016role} established a reference model of the ISPs roles in the anti-botnet life cycle from prevention to detection, notification, remediation, and recovery. 
They validated the model with a representative sample of Dutch ISPs and showed that ISPs spent most effort on the prevention and notification but less so on other activities~\cite{pijpker2016role}. 
The OECD also reported various initiatives by the ISPs in the fight against botnets in Australia, Germany, Ireland, Japan, Korea, the Netherlands, the UK, and the US~\cite{oecdisp}.

A large part of the role of ISPs in keeping the IoT ecosystem secure is similar to their role in combating botnets; indeed, many botnets comprise IoT devices. 
A prime example of such an endeavour is the cleanup of Mirai in the Netherlands. Cetin et al~\cite{ccetin2019tell} examined the ISPs' uses of walled gardens on Mirai-infected IoT devices. Traditionally, this practice is used to quarantine and notify customers whose computers were infected by malware and turned botnet. They found that the use of walled garden remediated 92\% of the Mirai infections within 14 days, and outperformed the uses of email notification. Their findings provide compelling evidence of the prominent role that ISPs play in keeping IoT secure, particularly in the after-fact events (i.e., after customers' IoT devices were infected by malware).

ISPs can also play an important role in preventing the spreading of customers' infected IoT hosted in their network.
One approach is for the ISPs to scan for vulnerable IoT devices and isolate them from the Internet before they are compromised~\cite{dietz2018iot}.
Another approach is a wide scan of vulnerable or infected IoT devices by central government agencies and asked the ISPs to notify the owner of these vulnerable/infected devices and ask them to take actions to remediate.
This approach is being undertaken in Japan, under the ongoing five-year NOTICE initiative~\cite{notice}.
In NOTICE, the National Institute of Information and Communications Technology (NICT) identifies vulnerable or compromised IoT devices; participating ISPs are informed and assume the responsibility of identifying and notifying their customers who own the devices.
While the effectiveness of this approach is yet to be evaluated, one key lesson learned from the previous Japanese government's effort to clean up malware and botnet on end users' computers (under the CCC and ACTIVE initiative, see Appendix~\ref{AppendixBackground}) suggested that participating ISPs in NOTICE are likely to lack the incentives to actively involve and invest in the effort.

The lack of incentives also deters ISPs from keeping IoT secure in general~\cite{cetin2017make}.
Therefore, working together with the ISPs to identify their attitudes in securing IoT, the barriers they face, and identify incentives to encourage them to overcome these obstacles is imperative.

\subsection{The role of individuals}
Individuals who own and use IoT devices are another key stakeholder in keeping IoT secure. 
In an ideal world, they would configure appropriate security and privacy settings and maintain good security hygiene until these devices reached their end of life. 
However, the reality is rather different. 

The field of human factors in security has long posited that security is a secondary, not primary, task~\cite{west2008psychology, sasse2005usable,zheng_presenting_2022}. Simply put, security tasks such as changing the default password on IoT devices are less important than the primary task: getting the new device up and running.
Security and privacy is not the most important factor in IoT device purchase behaviour either: Emami-Naeini et al~\cite{emami2019exploring} showed that they were ranked below features and price.

Individuals also have different levels of experience and skill when it comes to fixing compromised devices. A study of ISP customers asked to identify and remediate compromised devices on their home networks found that, while the participants were motivated, many could not complete all the recommended steps \cite{bouwmeester2021thing}.
Relying less on individuals to secure IoT and more on device makers to make IoT devices secure by design is a better solution. A study investigating responsibilities for smart device security and privacy found that users largely viewed the responsibility for security as shared between themselves and device manufacturers \cite{haney2021company}.

Although progress has been made and governments some countries (e.g. the UK~\cite{ukcodeofpractice}, Australia~\cite{aucodeofpractice}, and Japan~\cite{iotnewrule}) have mandated or recommended device makers to do so, there are still a large number of devices that are not secure~\cite{alrawi2019sok}. Therefore, individuals still have a role to play in securing the IoT ecosystem.

\section{Scope and methodology}
\label{Methodology}
In this study, we examine two key stakeholders in the IoT ecosystem: ISPs and individuals. For ISPs, we sought their views on their customers' IoT devices (hosted in the ISPs' networks), not the IoT devices used by the ISPs. For individuals, the scope was limited to ordinary users of IoT and in personal context only (i.e., in the household).

We aimed to answer three research questions:\\
\noindent Q1---\textit{What are ISPs' and individuals' attitudes towards the security and privacy of IoT?}
 
 Attitudes toward the security and privacy of IoT are fundamental to the ISPs' and individuals' course of action on securing IoT. Understanding their attitudes is key to why they do or do not keep IoT secure. Specifically in this study, we sought to understand their concerns about IoT, perceived likelihood that IoT can be compromised, their commitment to keep IoT secure, and their perception of the contribution made by other stakeholders to secure IoT.

\noindent Q2---\textit{What are the barriers that prevent ISPs and individuals to keep IoT secure?}

 For individuals, we sought to identify the barriers that prevent them from securing or remediating IoT devices.
 For ISPs, we focused on identifying the barriers that prevent them from investing and/or committing to keeping IoT devices on their networks secure.
 
\noindent Q3---\textit{What are the incentives to encourage ISPs and individuals to keep IoT secure?}

 Incentives are key to encourage ISPs and individuals to keep IoT secure. 
 We sought to identify internal and external incentives. We defined internal incentives as those that need to be applied to ISPs and individuals to motivate them to secure IoT. External incentives, on the other hand, are those that need to be applied to other relevant stakeholders (to encourage or make it easier for ISPs and individuals to keep IoT secure).
 
\subsection{Methodology}

\begin{figure}[htb!]
 \centering 
 \includegraphics[width=0.8\columnwidth,trim=0cm 0cm 4cm 0cm, clip]{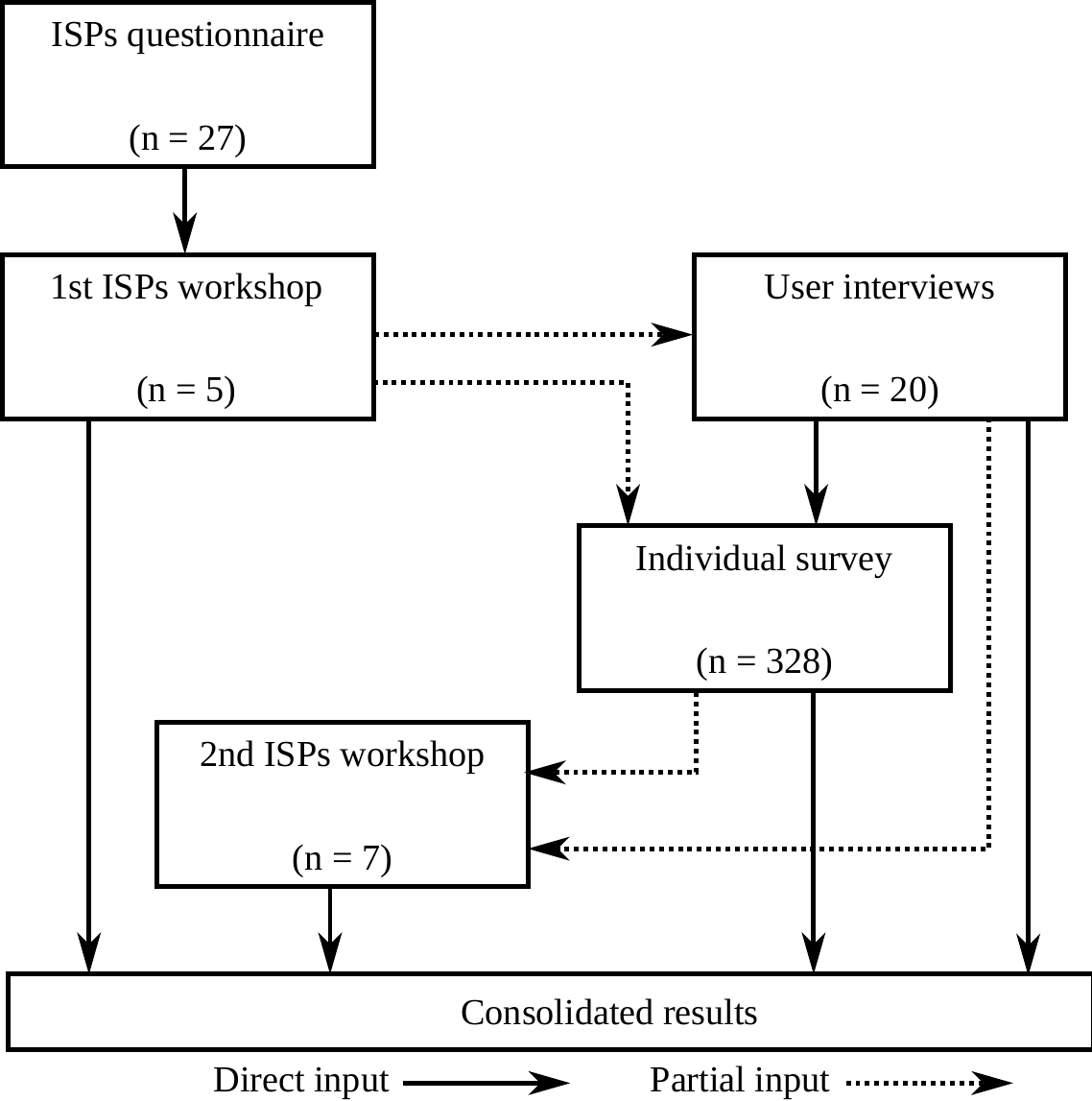}
 \caption{Overview of the methodology}
 \label{fig:method}
\end{figure}

We divided the research into studies of both ISPs and individuals. Based on the complex interactions between the stakeholders when trying to secure the IOT, we decided to adapt an iterative participatory action research based approach for our research~\cite{ozanne2008participatory}. Participatory action research has been successfully used in studying complex information systems~\cite{baskerville1999investigating,lazaroreasons}. The ISPs study consisted of a survey and two workshops. The individuals study consisted of user interviews and a survey. Each study is discussed in turn. An overview of the relationships between the different studies is shown in Fig.~\ref{fig:method}. The ISP survey was conducted initially, and the first ISP workshop was held after this to get feedback and clarification about survey findings. Both the survey and workshop then fed into the design of the individuals interviews and subsequent survey. Finally, a final ISP workshop was held to discuss the findings from the earlier studies.

\subsubsection{ISPs study} 

\paragraph{Recruitment and demographic of ISPs}

We advertised and recruited ISPs through the ICT Information Sharing and Analysis Centre Japan (ICT-ISAC)\footnote{Established in 2016, the centre is responsible for promoting a safe ICT society in Japan. Its members consist of organisations related to ICT, e.g. ISPs, security companies, and equipment manufacturers. \url{www.ict-isac.jp}}. ICT-ISAC regularly holds information exchange opportunities among ISPs, and participating ISPs are incentivized to exchange information on the status of security measures, issues, and knowledge of other ISPs. We requested that the ideal representative from the ISPs be in a position that can influence new policies or strategies in their organisation and have a sound understanding of the roles, responsibilities, and abilities of ISPs in keeping the Internet secure.

A total of 27 ISPs took part in the survey, 5 in the 1\st workshop, and 7 in the 2\nd workshop (4 of them also took part in the 1\st workshop). 
Participating ISPs encompassed mixed demographics with varied customers sizes, service coverage, and experiences with previous and current government's initiatives to promote safe ICT (Table~\ref{table:demographicISP}). Representatives from various government were also present, although primarily in an observatory role.
All ISPs participated voluntarily and no monetary reward was given.

\paragraph{ISP survey}
\label{methodpreworkshopsurvey}
The online (LimeSurvey) survey was designed to collect data about the ISPs' attitudes, barriers, and incentives to keep IoT secure.
It took approximately one hour to complete. We tested the questions with experienced members of the ICT-ISAC before launching the survey to ensure that the questions were sound and the survey's length was appropriate.

The survey has three parts (shown in Appendix~\ref{preworkshopsurvey}). 
Part 1 asks about demographic details of the ISPs. Part 2 asks about the attitudes and barriers. Specifically, it asks about concerns about their networks hosting infected/vulnerable IoT devices, perceived likelihood of their networks being attacked, roles and responsibilities of ISPs in securing IoT activities (and their current level of commitment to them), the priority of keeping the IoT ecosystem secure, and their perception of how other stakeholders have acted to secure IoT.

For barriers, we ask about the barriers which ISPs face internally within their organisation and externally---in terms of law and regulation, and in aiding NOTICE.
Part 3 asked about initiatives that ISPs viewed as incentives to motivate them to keep IoT secure in general and in aiding NOTICE.

We collected quantitative and qualitative data. Quantitative data were responses from rating, ranking, and multiple choices (single and multiple answers) questions. Qualitative data were from free-text questions. All questions asked for quantitative data, except for those about the barriers in NOTICE. 

For quantitative data, we performed statistical tests to determine the statistical significance of the findings. The Shapiro-Wilk test for normality was done first to determine whether the data being analyzed was normally distributed. For rating questions, the mean ($\mu$) was used to describe the central tendency of the responses if data were normally distributed and the median ($\tilde{X}$) and the mode ($Mo$) were used if data were not normally distributed~\cite{sullivan2013analyzing}. McDonald's Omega was also used as a measure for internal consistency\footnote{Cronbach’s Alpha does not suit small sample size ($n<30$)} and was found to be reliable ($\omega_{min}=0.72$).
For ranking questions, the Friedman test was used to determine the significance of the ranking scores. For multiple choices questions, the proportion test was used to determine whether the observed proportions of the selected answer choices were statistically significantly different. To compare responses between multiple sub-groups (mainly ISP sizes), we performed Mann-Whitney U tests. Where necessary, the significance boundaries of the statistical tests were Bonferroni adjusted to account for multiple comparisons. 

The small number of qualitative responses from the survey were thematically grouped by two researchers to identify common barriers that ISPs faced in the NOTICE project. This grouping was reviewed by other team members and iteratively refined until consensus was reached. The grouping can be seen in Table~\ref{tab:ISPCoding}.

\paragraph{ISP workshops}
We held two ISP online workshops (due to COVID-19 travel restriction) at two different stages in the study (Figure~\ref{fig:method}). Each of them served a different purpose. 

The 1\st workshop took place after the ISP survey. The aim was to obtain feedback and clarification about the findings from the survey. Feedback relevant to individuals was also used to inform the design of the individuals' study.
In the workshop, one researcher chaired the session, two took detailed notes and facilitated when required, and one of them presented a summary of key points at the end of the workshop.

The 2\nd workshop took place after the individuals' study. The objectives were to share preliminary findings from the individuals study with ISPs and seek their feedback where relevant. This was run in the same way as the 1\st workshop.

\subsubsection{Individuals study}
\label{sec:method:individualsStudy}

\paragraph{Recruitment and demographic of participants}
We advertised and recruited participants in Japan for both user interviews and survey through NTT Com Research\footnote{The company provides various data collection services in Japan, including Internet surveys, and has about 2.17 million registered users from all over the country as of March 2021 (\url{https://research.nttcoms.com}).}. Eligible participants must be living in Japan, are at least 18 years old, and were required to possess one IoT device besides a smartphone, tablet, or Wi-Fi router. 

A total of 20 participants took part in the user interviews and 328 in the survey. Participants encompassed a mixed demographic: gender, age, and IoT devices they have (Table~\ref{table:demographicIndividual}). 

\paragraph{User interviews}
The interviews aimed to gain a preliminary understanding of individuals' attitudes, barriers, and incentives to keep IoT secure. Findings from the interviews were primarily used to inform the design of the survey but some of the findings were also used to interpret the findings from the survey where appropriate.

A one-to-one, semi-structured one-hour online interview took place with each participant. The interview consisted of three parts (see Appendix~\ref{interviewscript} for the interview guide). Part 1 asked about the uses of IoT devices and factors affecting the uptake decisions. Part 2 asked about the security and privacy of IoT. This included the extent and the nature of concerns and supporting rationale, priority to keep IoT secure, the security hygiene practice on IoT devices and any problems they encountered, the perceived likelihood of their IoT devices being compromised and the rationale supporting their perception, and their experiences of remediating compromised IoT devices. Part 3 asked about possible initiatives that individuals viewed as incentives to motivate them to keep IoT secure. 

Interviews were recorded and manually transcribed. Each participant was then assigned a non-identifiable ID (P1--P20). We followed an established methodology for iterative thematic analysis \cite{braun2006using,morgan_iterative_2020}. After familiarising ourselves with the transcripts, two researchers generated initial codes that represented reoccurring aspects in the data. These codes (and their associated quotations) were discussed and defined with the larger researcher team, before repeating the coding process on all of the transcripts. After four rounds of this process the codebook was stable: we reached consensus on the meaning of the codes and their occurrences in the transcripts. At this point codes were grouped into themes. This process was supported by a review of prior literature, allowing us to identify reoccurring themes as well as new, divergent ideas in our data. These resulting themes were used to inform the survey structure and answer options, and the findings are discussed in Sections~\ref{sec:results-individuals} and~\ref{Discussion}. The codebook and coding matrix can be seen in Table~\ref{tab:userInterviewCoding}.

\paragraph{Survey}
The design of the survey is based on the findings from the interviews. It has three parts (see Appendix~\ref{consumersurvey} for an English translation) and follows the same topics and subjects as the interview. The original surveys uses a number of common Likert scales which have been translated with matching English ones.

In examining incentives, we also examined participants' willingness to pay (WTP) for IoT security services.
A contingent valuation (CV) method was used by asking participants to specify the amount they were willing to pay for four IoT security services: traffic monitoring, remote assistance, home visit, and a bundled service including all of these. CV is used widely in economic studies to determine the \textit{stated preference} of the WTP for a product or service (e.g.~\cite{blythe2020security, cohen2004willingness, seip1992willingness}). 

We applied the same statistical tests used in the ISP survey (Section~\ref{methodpreworkshopsurvey}) for the rating, ranking, and multiple-choice questions. Cronbach's alpha was used to determine the internal consistency of rating responses ($\alpha_{min}= 0.94$, suggesting that they were reliable). 

The online survey took about 20-25 minutes to complete. Prior to its launch, we iterated on the questions with four researchers.

\subsection{Ethics consideration}

The study was conducted after having been approved by the institutional review board. Permission was granted provided that we informed participants about the study, obtain their consent before data collection, and complied with Japan's Act on the Protection of Personal Information (APPI).
Individual participants gave informed consent to participate in the study, and no personally identifying information was collected. The ISP participants were aware of the time commitments, and results of this research are provided to the ISPs, which is an incentive for them. Each individual participant received \textyen5500 and \textyen1000 yen to participate in the interview and online survey, respectively.

\section{Results: ISPs}
\label{ResultsISPs}

We begin with the results from the ISPs' perspective. ISPs with more than 10,000 and less than 1 million customers were classed as `medium', with `small' and `large' ISPs on either side of this range. See Table~\ref{table:demographicISP} for full ISP demographics.

\subsection{Attitudes toward IoT security and privacy} \label{ispattitude}
ISPs had varying concerns about the potential impact of IoT security risks, the perceived likelihood that the risks will become materialized, and their priority to securing IoT ecosystem. 

\subsubsection{Nature of concern}
\label{ispconcern}
Overall, ISPs in Japan are most concerned about service disruption to customers as a result of their network hosting infected/vulnerable IoT devices (1\st), followed by the need to take social responsibilities as a result of IoT attacks occurred in their network (2\nd), and reputation damage (3\rd).
The loss of customers and market share, incurred financial cost and warning from relevant authorities are less of the concern, ranking as the bottom three. The ranking varies slightly by the size of ISPs. While service disruption ranks first for large and medium ISPs, social responsibilities ranks first for small ISPs.
Incurring financial cost also ranks highly (3\rd) in small ISPs, suggesting they are constrained by financial resources.
The full ranking can be found in Table~\ref{table:ISPconcernranking} in the Appendix. 

The finding that service disruption to customers ranks first makes logical sense as it can be considered a trigger of all other consequences: it can lead to the need to take social responsibilities and reputation damage.

The finding that social responsibilities ranked highly is rather intriguing.
It is likely to have a deep root in the longstanding corporate culture in Japan called \enquote{kyosei}, the concept of living and working together for the common good which emphasised taking care of the society and being part of the community~\cite{kaku1997path}.

Feedback from the 1\st ISP workshop shed some light on why ISPs felt strongly about social responsibilities.
One small ISP rationalised that the lack of known information about the potential impact on customers as a result of vulnerable IoT devices present in their network made them feel that they need to take more social responsibility than customers would [ISP5]. Another ISP, large in size, added that to them taking social responsibilities was also about being responsible for customer service delivery and for any inconvenience they may have caused to other ISPs if their network were infected/attacked by IoT botnet such as Mirai [ISP2].

\subsubsection{IoT attacks on ISPs' network} 
\label{isplikelihood}
Overall, most ISPs felt indecisive about the likelihood of the IoT attacks happening, either to or from their networks to other ISPs. On the scale of \enquote{mostly unlikely} to \enquote{unlikely}, \enquote{neutral}, \enquote{likely}, and \enquote{mostly likely}, the median ($\tilde{X}$) and mode ($Mo$) are \enquote{neutral}. The same $\tilde{X}$ and $Mo$ were also observed in small, medium, and large ISPs.

Feedback from the 1\st ISP workshop helped us to understand some of the rationale supporting the ISPs' ratings. One ISP rating \enquote{unlikely} based their rationale on past network statistics which have been pretty stable [ISP4]. Another, also rating \enquote{unlikely} was confident in their ability to control traffic volume and so did not think the attacks would be likely [ISP5]. One of the largest ISPs in Japan simply admitted not having enough information to make a prediction [ISP1].

\subsubsection{Priority, perceived responsibilities, and commitment} \label{isppriority}
Overall, most ISPs rated the priority for keeping IoT secure as \enquote{medium} (desired task) ($\tilde{X} = Mo$). The rating for large and medium ISPs were also \enquote{medium} but for small ISPs was \enquote{low} (tasked to be done if there is room) ($\tilde{X} = Mo$, with $p=0.016$ when comparing small against medium ISP, $U=10.5$).

\begin{figure}[tp]
 \centering 
 \includegraphics[width=\columnwidth]{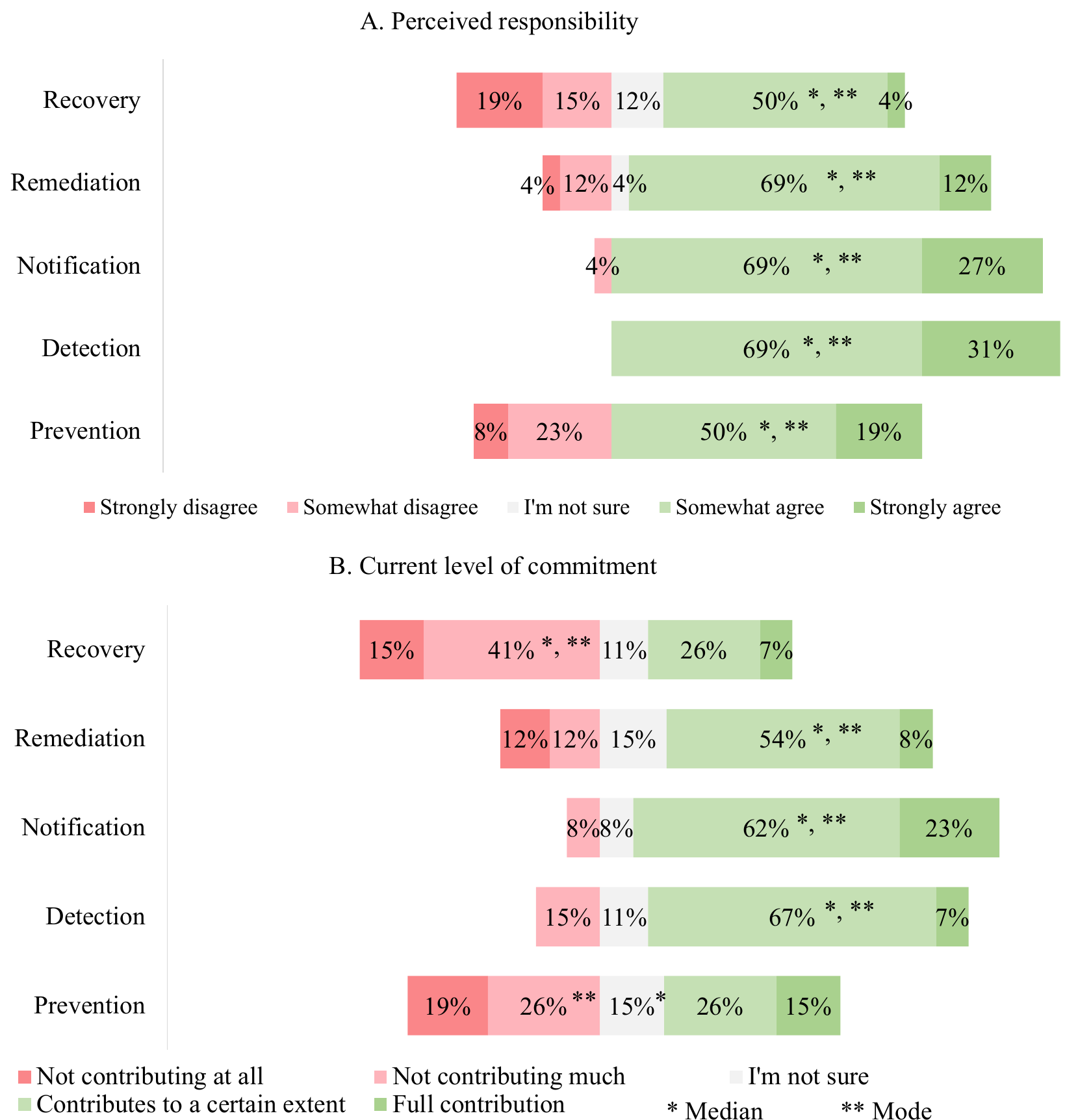}
 \caption{ISPs' perceived responsibilities on key activities to keep IoT secure and their current level of commitment}
 \label{fig:ISPcommitment}
\end{figure}

In examining the perceived responsibilities of ISPs on key activities to keep IoT secure, most ISPs agreed that they are responsible for all five key activities: prevention, detection, notification, remediation, and recovery ($\tilde{X} = Mo=$ \enquote{somewhat agree}) (Fig.~\ref{fig:ISPcommitment}). By ISP size, the same rating was also observed for most of the key activities, although there were a number of statistically significant variations: large ISPs view detection as less of a duty than medium ones, who in turn view it as less of a duty than small ones ($U=40$ and $U=21$, both $p<0.05$). Large ISPs also viewed remediation actions as less of their responsibility than medium and small ones ($U=43.5$, $p<0.05$).

Their current commitment to these five activities was fairly consistent with their perceived responsibilities.
Overall, most ISPs made some contributions to all these activities ($\tilde{X} = Mo =$ \enquote{contributed to a certain extent}), except for prevention activities ($\tilde{X}= $ \enquote{not sure}, $Mo=$ \enquote{Not contributing much}) and recovery activities ($\tilde{X} = Mo =$ \enquote{Not contributing much}).

These findings suggested that although ISPs felt that they were responsible for key activities in keeping IoT secure. One medium ISP explained in the 1\st ISP workshop that part of the reasons was the view that the government's NOTICE initiative would reduce the cost associated with the recovery activities because NOTICE would reduce the likelihood of IoT attacks from happening [ISP3]. 
Like large and medium ISPs, small ISPs also perceived the likelihood of IoT attacks as medium (Section~\ref{isplikelihood}). However, unlike the two counterparts, small ISPs rated the priority of keeping IoT secure as low. They were also unsure about their commitment to half of the key activities to keep IoT secure. One possible root cause is the lack of knowledge and information to make informed decisions about investing in keeping IoT secure, discussed below.

\subsection{Obstacles to keeping IoT secure} \label{ISPobstacles}

\subsubsection{Internal barriers}\label{ISPobstaclesinternal}
Overall, a shortage of skilled staff to keep IoT secure was chosen most (human resource, 44\%) (Table~\ref{table:ISPbarrier}). About four in ten ISPs identified convincing executives in the organisation to buy into the idea of securing IoT as a barrier (executive buy-in, 44\%). Budget constraints (financial, 37\%) and lack of technology without new investment (technology, 37\%) were also identified but by fewer ISPs.
These barriers varied considerably by ISP size. 
The distribution of internal barriers were statistically significantly different from chance for the overall data as well as large ISPs and medium ISPs (see Table~\ref{table:ISPbarrier}). 

Feedback from the 1\st ISP workshop clarified the nature of these barriers.
One small ISP explained that although the financial budget is not the barrier itself, allocating the budget to secure IoT is hard because it is difficult to determine the effort needed [ISP5]. The nature of this feedback suggested that the small ISP lacked the knowledge to make informed decisions about planning and investing in securing IoT. Regarding executive buy-in, one large ISP explained that trying to convince executives to see the point of keeping IoT secured and its benefits was more difficult than for other types of threats [ISP1]. For example, in the case of spamming, it was easier to explain the clear benefit of reducing the number of abuses which led to cost savings. Attempting to do the same in the context of IoT was difficult.

\subsubsection{External barriers}\label{ISPobstaclesexternal} 
\begin{figure}[!tp]
 \centering 
 \includegraphics[width=\columnwidth]{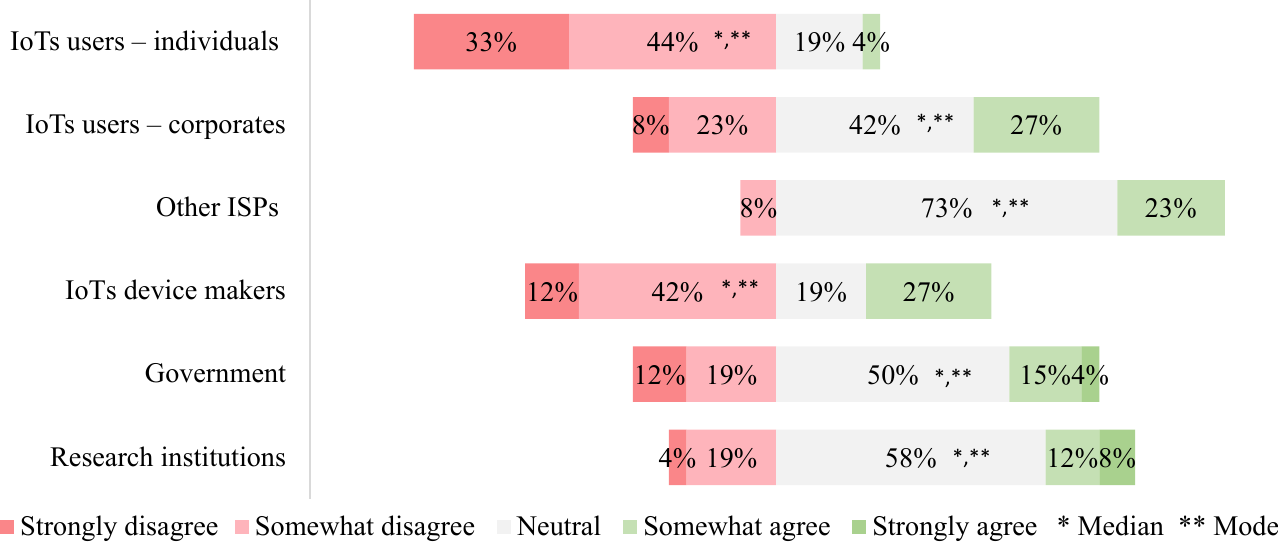}
 \caption{ISPs' views of how other stakeholders sufficiently keep IoT secure}
 \label{fig:ISPperceptionotherparty}
\end{figure}

To understand external barriers, we first examined ISPs' views on how other stakeholders have sufficiently keep IoT secure. 
Most ISPs had a mixed, but overall neutral opinion on how IoT corporate users, other ISPs, government, and research institutes have sufficiently done their part in keeping IoT secure ($\tilde{X}=Mo=$ \enquote{neutral}). 
However, they were less optimistic about individual IoT users and IoT device makers ($\tilde{X}=Mo=$ \enquote{somewhat disagree}) (Fig.~\ref{fig:ISPperceptionotherparty}). There were no statistically significant variations by ISP size.

Feedback from the ISP workshop shed light on why: 
ISPs felt that individual users were less likely to manage the devices properly and more likely to have problems with IoT security because these users make careless configuration settings on routers.
These points agreed with individual users' perceptions: lack of knowledge and skills to keep IoT secure was one of the barriers that individuals reported they faced (Section~\ref{individualbarrier}). 
Regarding IoT device makers, even though the new regulation in Japan mandates all IoT devices to be secure by default, 
there were a large number of devices released to the market before the new regulation came into effect, which remain a problem.

In examining regulatory barriers, Telecommunication Business Law Article 4 (Protection of \enquote{secret of communication}) was the biggest barrier ($\tilde{X}=$ \enquote{somewhat agree}, $Mo= $ \enquote{strongly agree}). In a nutshell, this law does not allow ISPs to monitor/inspect customers' Internet traffic, even if some of them are malicious, and prohibits ISPs in Japan from implementing a walled garden even though Cetin et al~\cite{ccetin2019cleaning} showed that it was an effective practice to mitigate the risks from IoT botnets. ISPs viewed other law and regulation less of a barrier (Japan APPI) ($\tilde{X}=$ \enquote{neutral}, $Mo= $ \enquote{somewhat disagree}), the EU General Data Protection Regulation (GDPR) ($\tilde{X}=Mo=$ \enquote{neutral}), and Net neutrality ($\tilde{X}=Mo=$ \enquote{neutral}).

\subsubsection{Barriers to notification}
\label{ispnoticebarrier}

These barriers were identified from the qualitative responses in the ISP survey and then discussed in the workshops. In identifying customers whose IoT devices were identified as vulnerable/infected, two issues emerged. First, the dynamic nature of IP addresses made it difficult, or often not possible, to match the list of IP addresses received from NOTICE and the customer's IP address. 
Second, traffic logs were not always available. Even if they are available, the sheer volume of logs made the race to find a needle in the haystack against time very challenging. 

In notifying customers whom ISPs identified as the \enquote{real} owner of the IP addresses behind the infected/vulnerable devices, there were three issues. First, ISPs did not always have customers' up-to-date contact details, especially their personal e-mail addresses. Currently, ISPs notified customers via carrier email, an email that ISPs created for customers when they signed up, but most customers rarely (or never) check it. Feedback from the ISP workshop indicated that this issue is a major bottleneck. Some small ISPs tried to reach customers by phone or home visit but did not always succeed. 
Hence, most ISPs relied largely on carrier email to notify customers.

Second, in many cases, ISPs were not able to notify the owner of the infected/vulnerable devices. The owner of the device and the owner of the internet connection were not always the same. For example, it is not uncommon for parents to sign a contract with ISPs on behalf of their children, or for a landlord to sign the contract but the tenants own the IoT devices.

Finally, even when the ISPs managed to notify customers successfully, many customers did not understand or were not convinced about the risks this process aimed to mitigate. Many of them ignored the problem and did not remediate their devices. Feedback from the first ISP workshop shed light on the likely root causes. The risks from \enquote{vulnerable} devices is future risk; it is hard for customers to imagine how or if the risks will be materialised. The general public was also perceived by ISPs to lack awareness about the consequences of not fixing IoT devices, meaning that some customers might not recognise it as an issue or see the benefits from protecting or remediating these devices. 

\subsection{Enabling a secure IoT}
\label{ispincentive}

\subsubsection{Internal measures}
\label{ispincentiveinternal}
When asked which measures may help motivate their organisation to invest more in securing the IoT ecosystem, most chosen by ISPs was acquiring technology to detect infected and/or vulnerable IoT devices (50\%), followed by executive and staff buy-in to the idea of securing IoT (33\%) (Table~\ref{table:ISPincentives}).
This is in contrast with the ranking of barriers (Table~\ref{table:ISPbarrier}), where executive buy-in was ranked higher than a lack of technology. 

Only a few ISPs (12\%) chose monetizing IoT (such as by offering IoT security services to customers), and none wanted to transfer costs associated with keeping IoT secure to customers. 
In the workshop, the ISPs expanded on this. The first concern was that an extremely diverse range of IoT devices made it difficult to design a security service package that covers all or a reasonable fraction of IoT devices. Further, the ISPs were afraid of a large gap between the service that ISPs can offer and that customers expect from ISPs, and of over-promising to customers if the ISPs fail to detect and block IoT malware.
Moreover, convincing executives in the organisation to approve any new services is usually hard in general, even harder for a security service because it is not ISPs' core service.

\begin{table}[htb!]
\caption{\% of ISPs per general incentives to secure IoT}
\centering
\scriptsize
\setlength{\tabcolsep}{4pt}
\begin{tabular}[p]{p{3.6cm}p{0.25cm}p{0.25cm}p{0.25cm}p{0.25cm}p{0.25cm}p{0.25cm}p{0.25cm}p{0.25cm}}
 \toprule
   &\multicolumn{2}{c}{Overall} &\multicolumn{2}{c}{Large} &\multicolumn{2}{c}{Medium} &\multicolumn{2}{c}{Small}\\
Incentives &$n$ &\% &$n$ &\% &$n$ &\% &$n$ &\%\\
  \midrule
\multicolumn{9}{l}{Internal}\\
\midrule
Tech to detect infected IoT &13&50&4&50&7&50&2&50\\
Executive \& staff buy-in &9&33&4&50&4&27&1&25\\
Offer IoT security service &3&12&2&25&1&7&0&0\\
Transfer cost to customer &0&0&0&0&0&0&0&0\\ 
\midrule
\multicolumn{9}{l}{External}\\
\midrule
(M) Device makers secure IoT  &19&70&7&88&8&57&4&100\\
(G) Amend law to restrict traffic &15&58&5&63&9&60&1&25\\
(G) Public recognition for ISPs &14&52&4&50&7&50&3&75\\
(G) Government subsidy &13&50&4&50&8&53&1&25\\
(U) Individuals keep IoT secure &9&33&1&13&6&40&2&50\\
(U) Corporate keep IoT secure &9&33&1&13&6&43&2&50\\
(G) Amend law to monitor traffic &8&31&1&13&7&50&0&0\\
(G) Public-private sector partnership &6&22&3&38&2&13&1&25\\
(G) Make net neutrality optional &3&11&1&13&1&7&1&25\\
(R) Input from research community &3&12&0&0&2&13&1&25\\
(I) Equal responsibilities for ISPs &1&4&0&0&1&7&0&0\\
(I) ISPs performance dashboard &1&4&0&0&1&7&0&0\\
\midrule
Other &1&4&1&13&0&0&0&0\\
Not sure &0&0&0&0&0&0&0&0\\
\midrule
$\chi^2(17)$ &\multicolumn{2}{p{0.9cm}}{115.70**} &\multicolumn{2}{p{0.85cm}}{48.76**} &\multicolumn{2}{p{0.85cm}}{66.28**} &\multicolumn{2}{p{0.85cm}}{29.53*}\\
 \bottomrule
\end{tabular}
\label{table:ISPincentives} \vspace{0.3em}
\parbox{\columnwidth}{ \scriptsize
(M) = device maker, (U) = end users, (G) = government, (I) = ISPs community, (R) = research community,  *, ** $p<0.05$, and $p<0.001$, respectively.
}
\vspace{-2em}
\end{table}

\subsubsection{External measures} 
\label{ispincentiveexternal}
A large proportion of ISPs wanted device makers to secure IoT devices (70\%) (Table~\ref{table:ISPincentives}).
This is consistent with ISPs views on obstacles (Section~\ref{ISPobstaclesexternal}). 
Fewer ISPs wanted users, individual or corporate, to secure IoT devices (33\%), despite ISPs stating that end users have insufficiently kept IoT secure (Section~\ref{ISPobstaclesexternal}). 

Some incentives from governments appealed to ISPs, with \enquote{amending the law to allow ISPs to restrict customer traffic} being ranked highest (58\%).
Essentially, this would allow ISPs to block malicious IoT traffic and set up a walled garden.
About half of ISPs also viewed public recognition of ISPs for keeping IoT secure as an incentive (58\%), which may be related to the \enquote{kyosei} principle.
Government subsidy also appealed to half of ISPs, suggesting that in the context of securing IoT in general, one in two ISPs would rather seek financial support from the government than making money from or transfer costs to customers.
Some ISPs (22\%) considered a public-private sector partnership such as NOTICE an incentive.

\subsubsection{Notification measures}
\label{ispincentivenotice}

\begin{table}[htb!]
\caption{\% of ISPs per incentives for NOTICE}
\centering
\scriptsize
\setlength{\tabcolsep}{4pt}
\begin{tabular}[p]{p{3.6cm}p{0.25cm}p{0.25cm}p{0.25cm}p{0.25cm}p{0.25cm}p{0.25cm}p{0.25cm}p{0.25cm}}
 \toprule
   &\multicolumn{2}{c}{Overall} &\multicolumn{2}{c}{Large} &\multicolumn{2}{c}{Medium} &\multicolumn{2}{c}{Small}\\
Incentives &$n$ &\% &$n$ &\% &$n$ &\% &$n$ &\%\\
  \midrule
(M) Public awareness of risks &18&69&5&63&11&79&2&50\\
(R) Effective notification method &10&38&3&38&6&43&1&25\\
(N) Tech to authenticate notification &9&33&1&13&6&40&2&50\\
(N) Tech to track notification &9&33&4&50&5&33&0&0\\
(R) Support from academia to identify effective notification method &8&31&0&0&6&43&2&50\\
(G) Government subsidy &7&27&4&50&2&14&1&25\\
(N) More data from NOTICE to identify owners of vulnerable devices &4&15&4&50&2&13&1&25\\
\midrule
$\chi^2(7)$ &\multicolumn{2}{c}{31.42**} &\multicolumn{2}{c}{15.81*} &\multicolumn{2}{c}{24.01**} &\multicolumn{2}{c}{6.03}\\
 \bottomrule
\end{tabular}
\label{table:ISPincentivesnotice} \vspace{0.3em}
\parbox{\columnwidth}{ \scriptsize
(M) = non specified party, (G) = government, (R) = research community, (N) = NOTICE project, *, ** $p<0.05$, and $p<0.001$, respectively.}
\vspace{-1em}
\end{table}

With the NOTICE initiative, the government detects vulnerable and compromised IoT devices, and passes a list of IP addresses to ISPs. To help with this, first and foremost ISPs wanted the risks and impacts of leaving IoT devices vulnerable/infected known to the public so that the users understood the importance of remediating these devices (69\%) (Table~\ref{table:ISPincentivesnotice}).
This made sense given that ISPs identified users ignoring the problem when the ISPs notified them about vulnerable/infected IoT devices as a barrier (Section~\ref{ispnoticebarrier}).

The next most selected incentive was published data on effective notification methods to notify customers about the infected/vulnerable IoT devices (38\%), most likely due to ISPs identifying the use of carrier email as a bottleneck in the notification process.
About a third wanted technology that allows the notified customers to authenticate the notification (to ensure it is from the ISP and not from a scammer or spammer) (33\%), the ISPs to track the status of notifications sent to the customers (33\%), and support from the research community to identify effective channels to notify customers (31\%), the latter of which we examined in this study (Section~\ref{indivualincentiveinternal}).

\section{Results: Individuals}
\label{sec:results-individuals}

Now we consider the individuals' perspective. These results are based on our analysis of 20 interviews and 328 survey participants as described in Section~\ref{sec:method:individualsStudy}. Throughout this section we use quotations from the interviews to help contextualise the qualitative results of the survey. 

\subsection{Attitudes toward IoT security and privacy}

\subsubsection{Concerns by types of IoT} \label{concerndevice}
Overall, participants mostly feel unconcerned about the security and privacy of the different types of IoT. 
The $\tilde{X}$ and $Mo$ are \enquote{not concerned} or \enquote{Neutral} for most devices excepted for smartphone/tablets, smart security devices, smart gadgets for kids and home assistants ($\tilde{X} = Mo= $ \enquote{somewhat worry}).

The observed \enquote{somewhat not worry} on smart health devices was puzzling; these devices tend to collect and process data about the users' health. The finding suggested that participants may not view this data as sensitive enough to worry about.

\begin{figure}[htb!]
 \centering 
 \includegraphics[width=\columnwidth]{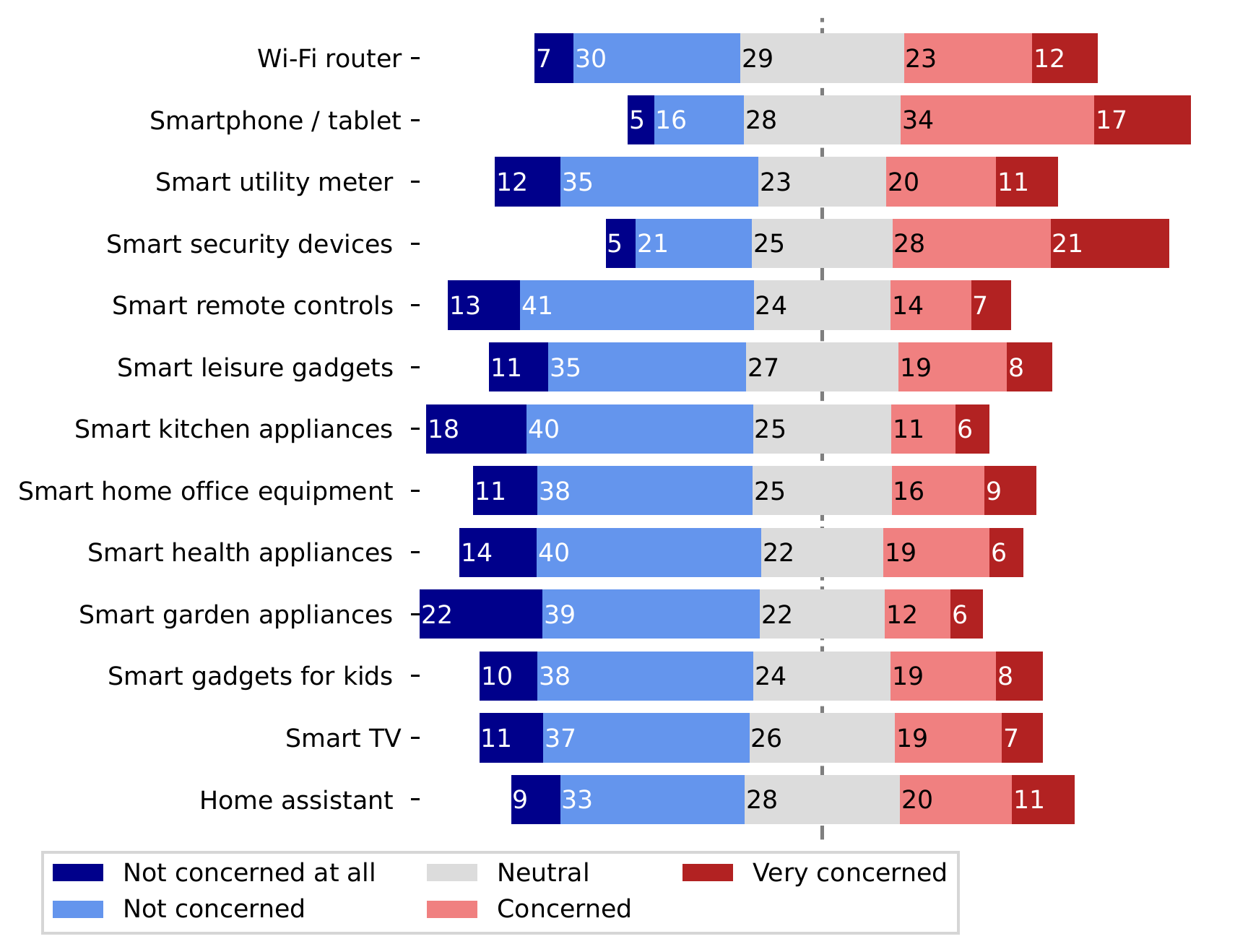}
 \caption{Proportion of participants and their concerns about security and privacy of IoT by types of IoT.}
 \label{fig:consumerconcerndevice}
\end{figure}

\subsubsection{Nature of concerns}
Participants in the survey were most concerned about the financial loss (e.g. leakage of credit card details linked to IoT devices), unsurprising given the importance of financial resources.
Invasion of privacy (e.g. data leakage from smart CCTV or TV) ranked second, and participants ranked inconvenience of disruption to daily activities (e.g. being locked out from home as a result of unauthorised locking of the door) third.
Physical effect, and harms to household assets were of the lowest concern.
There were virtually no differences between the age groups and genders throughout this analysis. All rankings are statistically significant ($\chi^2(5)$ at $p<0.001$). 

\subsubsection{Perceived likelihood of devices being compromised} \label{perceivedlikelihood}
Participants are uncertain about the likelihood of their IoT devices being compromised.
Those that consider it unlikely that their devices had been compromised have varying reasons for believing so. The financial angle was predominant, with 53\% selecting \enquote{no suspicious activities on financial services}. It suggested that participants were protective of financial resources and resort to them as a means to determine the likelihood of IoT risks becoming materialised.

These options highlight incorrect mental models about the activity of IoT malware and anti-virus as if it is a panacea to all cyber threats.
In the user interviews, 9 participants ($9/20$, $45\%$) also turned to anti-virus report to determine whether their IoT devices have been compromised; as an example, P13 explained, \textit{\enquote{Since the monthly report of the security software shows that there have been zero detection so far, I believe that I am safe even to the extent that I am not fully aware of in my PC. Based on this, I assume that IoT devices connected to the same environment are also safe.}}
Cetin et al~\cite{ccetin2019cleaning} observed a similar mental model concerning anti-virus and IoT among users in the Netherlands; their participants ran anti-virus software on PCs to solve the infection of an IoT device. Theirs and ours findings suggested that, despite cultural differences between both countries, their residents have similar misconceptions about anti-virus and IoT.

\subsection{IoT security hygiene and barriers}
\label{individualbarrier}
Participants in the survey indicated configuring various security and privacy settings as part of IoT device setup.
The three most selected features were two-factor authentication (2FA, $38\%$), despite only one participant mentioning 2FA in the user interviews ($1/20$, $5\%$), setting up account and password to restrict access ($37\%$), and updating software and firmware ($36\%$).
Participants touched other features less: changing default password ($31\%$), setting mode of software update (to auto or manual) ($30\%$), deleting any old data that may still be on the device ($19\%$), and restricting data sharing ($15\%$).

\paragraph{Barriers to remediating compromised IoT}
\label{individualBarriersToRemediation}
We asked participants who suspected that their IoT devices may have been infected (\enquote{Maybe/I don't know}, \enquote{Probably}, and \enquote{Definitely}) ($n=169$) if they had tried to remediate the devices.
The majority of them did not ($74\%$).
The most frequently selected reason was \enquote{not knowing where to start} ($56/125$, $45\%$), followed by \enquote{easier to replaced it with new device} ($36/125$, $28\%$) and \enquote{fixing it did not worth the stress and time} ($36/125$, $28\%$). 

For those trying to remediate the devices ($44/169$, $26\%$), a small proportion did not experience any problem at all ($21\%$). These figures are in line with those found by Bouwmeester et al~\cite{bouwmeester2021thing}.
For those facing some difficulties, the main problem was too much technical knowledge and skill requiring them to understand ($20/44$, $46\%$), resonating with the earlier finding of the participants' lack of skill and knowledge to fix the devices.
The remaining reasons were \enquote{insufficient support from ISP} ($31\%$), \enquote{friends and family} ($30\%$), and \enquote{device makers} ($25\%$).

\subsection{Keeping IoT secure}
\label{individualincentive} 

\subsubsection{Individual triggers}
\label{indivualincentiveinternal}
We ask participants about triggers that would make them aware of IoT security and privacy or that would want them to keep their devices secure.
The largest proportion selected \enquote{watching/reading news/stories about past attacks on IoT devices} ($53\%$) followed by \enquote{direct experiences of IoT attacks} (47\%).
Subsequent triggers on \enquote{seeing advertisements that educate people about the dangers of IoT devices} ($37\%$), \enquote{hearing a close friend or family member talking about the dangers of IoT} ($34\%$), \enquote{being told about the dangers of IoT at school or work} ($23\%$), and \enquote{watching TV shows or celebrities talk about the dangers of IoT} ($20\%$). The proportion test determined that the observed proportions were statistically significant ($\chi^2=409.02$, $df=7$, $p<0.001$).

Findings from the interviews also helped us to understand how media exerted influence on participants.
From one participant: \enquote{\textit{Well, I am concerned about X [home assistant device]. I've heard on the Internet news that people have had their credit card details stolen, so I think we have to be very careful about security. I had never been concerned about security before, but when I heard about such security cases, I started to feel scared little by little} [P8].}

\subsubsection{Views of other stakeholders} \label{indivualincentiveexternal}
\begin{figure}[htb!]
 \centering 
 \includegraphics[width=\columnwidth]{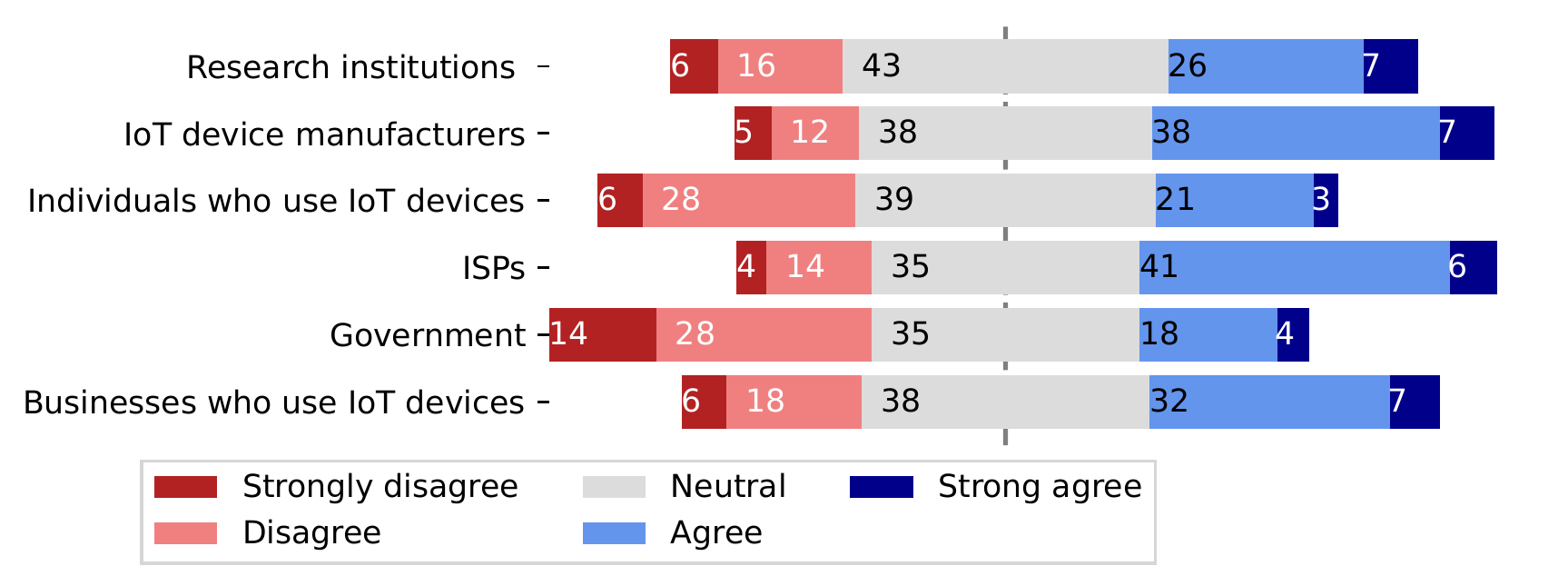}
 \caption{Participants' view on to what extent stakeholders have taken a sufficient role in keeping IoT secure.}
 \label{fig:consumerperceptionotherparty}
\end{figure}

We examine how individuals perceived how other stakeholders are doing their jobs in securing IoT.
Overall, participants in the survey feel ambivalent about how other stakeholders are taking sufficient responsibility in keeping IoT secure.
The ($\tilde{X}$) and ($Mo$) for the rating of all stakeholders are \enquote{Neutral} (Figure~\ref{fig:consumerperceptionotherparty}). 

\subsubsection{Consumer view on ISP initiatives}
\label{consumerViewOnInitiatives} 
Participants in the survey wanted ISPs to establish various initiatives to help customers keep IoT secure.
The most selected initiatives concerned \enquote{ISPs monitoring suspicious traffic to-and-from IoT devices}, \enquote{notifying customers before blocking} ($55\%$) and \enquote{automatically block the traffic} ($54\%$), demonstrating a strong desire for ISPs to take a proactive approach to keep network supporting IoT secure (Table~\ref{table:IndividualincentivesfromISPs}).

More than half of the participants in the user interviews ($13/20$, $65\%$) were also keen on having ISPs monitor IoT traffic.
For example \textit{\enquote{I am not familiar with the Internet, so if I were asked \enquote{is it okay to block malicious traffic?}, I would answer \enquote{If it is dangerous, yes, please}\textelp{} I don’t know how to confirm if the traffic is dangerous or not, so I am grateful if ISPs do that}} [P6].

\begin{table}[htbp!]
\caption{Participant support for ISP incentives}
\label{table:IndividualincentivesfromISPs}
\centering
\scriptsize
\setlength\tabcolsep{0.5em}
\begin{tabular}[p]{p{7cm}cc}
 \toprule
  Incentive &$n$ &$\%$ \\
 \midrule
  1. Monitor suspicious com. from IoT devices and \textit{notify} customers. &180 & 55  \\
  2. Monitor suspicious com. from IoT devices and \textit{automatically block} it. &177 & 54  \\
  3. Notify users about important software/firmware router updates. &140 &43 \\
  4. Set appropriate security settings on devices before shipping. &119 &36  \\
  5. Have a call center where users can easily consult about security. &114 &35  \\
  6. Be transparent about how ISPs protect users' communication and personal data. &114 &35  \\
  7. Provide online service to help users repair compromised devices. &103 &32 \\
  8. Provide home visit service to repair compromised devices. &72 &22 \\
  9. Educate users about the potential risks of using IoT. &71 &22  \\
 \bottomrule
\end{tabular}\vspace{0.3em}
\parbox{\columnwidth}{ \scriptsize
328 participants total. The observed proportions among the incentives were statistically significant ($x^2=344.21$, $df=9$, $p<0.001$).
}\vspace{-1em}
\end{table}

However, monitoring traffic and notify or block suspicious traffic faces two major challenges. 
First, the legal framework in Japan makes it difficult for ISPs to monitor traffic on a large scale.
The second challenge lies in the task of notifying customers; the difficulties in reaching customers is one major bottleneck for the ISPs participating in the NOTICE initiative (Section~\ref{ispnoticebarrier}).

\subsubsection{Willingness to pay (WTP) ISPs for IoT security}
\label{consumerWillingnessToPay} 

ISPs have an opportunity to offer services to consumers to help them secure their IoT devices.
Based on the interviews and further discussions, we explore four different potential services: Monitoring IoT and notifying consumers of suspicious activity, providing remote support for IoT devices, home visits for technical assistance, and a bundle services of all the previous options.
Table~\ref{tab:willingnesstoPay} summarises our analysis, breaking it down by age groups and gender.

\begin{table}[htb!]
    \caption{Willingness to pay ISPs for IoT Security services}
    \label{tab:willingnesstoPay}
    \centering
      \scriptsize
\setlength\tabcolsep{0.18em} %
\newlength\wtpspacing %
\setlength\wtpspacing{0.3pt} %
    \begin{tabular}{l@{\hspace{-2pt}}c@{\hspace{\wtpspacing}}cc@{\hspace{\wtpspacing}}c@{\hspace{\wtpspacing}}cc@{\hspace{\wtpspacing}}c@{\hspace{\wtpspacing}}cc@{\hspace{\wtpspacing}}c@{\hspace{\wtpspacing}}cc} \toprule
                 & \mcthree{Monitor \& notify} & \mcthree{Remote assistance} & \mcthree{Home visit}       & \mcthree{Bundle service}  \\ \midrule
                 & \%$\neq$0                   & $\mu$                       & $\sigma$                   & \%$\neq$0                  & $\mu$           & $\sigma$ & \%$\neq$0 & $\mu$           & $\sigma$ & \%$\neq$0 & $\mu$    & $\sigma$\\ \midrule
overall\ts{**}     & 69                          & \Ca 635                     & 1094                       & 61                         & \Ca 665\ts{**}  & 1135     & 58        & \Cb 1085\ts{**} & 1802     & 73        & \Cc 1226 & 1976    \\ \midrule
18--24\ts{**}  & 74                          & \Cb 1106\ts{*}              & 1617                       & 66                         & \Ca 931\ts{*}   & 1364     & 59        & \Cc 1220\ts{*}  & 1820     & 74        & \Cd 1259 & 1753    \\
25--34\ts{**}  & 67                          & \Ca 644                     & 1070                       & 56                         & \Ca 676         & 1180     & 53        & \Ca 682\ts{**}  & 1187     & 67        & \Cb 1044 & 1737    \\
35--60\ts{**}  & 64                          & \Ca 472                     & 818                        & 59                         & \Ca 569\ts{**}  & 958      & 60        & \Cb 1194\ts{*}  & 1855     & 70        & \Cc 1332 & 2179    \\
>60\ts{**}    & 72                          & \Ca 385\ts{*}               & 562                        & 65                         & \Cb 520\ts{**}  & 1008     & 61        & \Cc 1235\ts{*}  & 2140     & 81        & \Cd 1264 & 2172    \\
                 & \mcthree{$r=-0.16$\ts{**}}  & \mcthree{$r=-0.092$\ts{*}}  & \mcthree{$r=0.029$}        & \mcthree{$r=-0.0053$}     \\ \midrule
Male\ts{**}   & 68                          & \Cb 599                     & 1062                       & 58                         & \Ca 568\ts{**}  & 1004     & 53        & \Cb 992\ts{**}  & 1795     & 71        & \Cc 1175 & 1919    \\
Female\ts{**} & 70                          & \Ca 669                     & 1125                       & 64                         & \Ca 754\ts{**}  & 1241     & 63        & \Cb 1171\ts{**} & 1809     & 75        & \Cc 1273 & 2032    \\
\bottomrule   
\end{tabular} 
\vspace{0.3em}
\parbox{\columnwidth}{ \scriptsize
All numbers in Yen per month. \%$\neq$0 is the percentage of people willing to pay >0. The colors show the order of the mean ($\mu$) for each of the services. The smallest mean is given the lightest color, and the colors darken if there is a statistically significant difference between the current mean and the next largest mean, based on a one-sided paired Wilcoxon rank test. This statistical test is further shown by **/* at the value of the lower cell, showing $p<0.01$/$p<0.05$ confidence respectively. If more than one cell contains the same color, there is no statistical significant variation between the means for these services. For the analysis by age, Pearson's $r$ indicates the correlation between age and willingness to pay. As before, **/* indicates statistical significance with $p<0.01$/$p<0.05$.
}
\end{table}

Overall, almost all participants would be willing to pay for some service. 
WTP is decreasing with age in Japan; however, this appears to be 'U' shaped, where the 60+ group is consistently the least willing group.
Women are willing to pay statistically significantly more than men for most services. 

The proportion of customers willing to pay anything as well as the absolute amounts participants are willing to pay are promising. 
They support the findings from Rowe and Wood \cite{rowe2013home} who found customers in the US were willing to pay ISPs to improve cybersecurity.
However, feedback from the 2\nd Japan ISP workshop indicates that the amount that participants are willing to pay is too low for ISPs to cover the costs of providing the service (Section~\ref{ispincentiveexternal}). This perhaps then hints at a lack of commodification of IoT monitoring and remediation services in Japan.

\subsubsection{Consumer view on IoT device makers} 
\label{consumerIoTDeviceIncentives} 
Participants first and foremost want IoT device makers to make secure devices ($59\%$), supporting the findings from user interviews where half of the participants wanted \enquote{secure by default} IoT devices ($10/20$, $50\%$).
The second most selected choice was regular software/firmware updates ($53\%$) which is similar to securing IoT devices.

These top-ranked measures clearly indicate that consumers believe IoT security is the manufacture's responsibility. More user-centered incentives such as \enquote{providing a repair service}, \enquote{call center support} and \enquote{help in stores} rank bottom (32\%, 38\% and 31\%).

\subsubsection{Consumer view on government}
\label{consumerGovernmentIncentives} 
We explore with the participants several possible measures by governments to secure IoT. 
Here \enquote{changes to the law to allow ISPs to monitor and block suspicious Internet traffic} ranked top with 62\%. \enquote{More stringent IoT regulations} ranked 2\nd, \enquote{arresting cybercriminals} 3\rd, and \enquote{raising awareness about the dangers of IoT} ranked 4\th.

The proportion test determined that the observed proportions among the measures were statistically significant ($\chi^2=339.67$, $df=5$, $p<0.001$).

User interviews allowed us to understand some of the rationales for individuals' keen interest for the government to tighten the law and regulation around IoT; over half of participants want governments to undertake such initiatives.
For example: \textit{\enquote{I am concerned about regulation. If we don’t have any regulation, people do as they like, and personal information may be leaked, so I want them \textins{government} to make proper regulation}} [P5]. It should be noted, however, that users may not be aware of all of the implications of a new law and that these results may overestimate support for such changes as the questions were focused primarily on the benefits.

\section{Discussion: Lessons learned} 
\label{Discussion}
Our study provides several useful insights to various stakeholders keen to promote the safety of the IoT ecosystem. These insights are based on our analysis of our ISPs questionnaire, both ISPs workshops, the user interviews and the user survey. 

\subsection{Macro socio-technical challenges}

Our multi-stage methodology has made it obvious that individual actors are almost powerless to improve IoT security by themselves, as the following challenges demonstrate. 

\subsubsection{Keeping IoT secure is not a priority for ISPs and individuals} \label{lessonpriority}
The first important lesson is that despite ISPs and individuals' concerns about the security and privacy risks of IoT, they do not prioritise the need to secure IoT.
For individuals, this lesson supports the view from research in human factors in security which posits that security is not a primary but a secondary task~\cite{west2008psychology, sasse2005usable}.
For ISPs, their core business is providing Internet service and so it is understandable that securing the IoT ecosystem is not their priority, although doing so has its merits.
This lesson is crucial; it highlights the importance of implementing appropriate incentives to encourage ISPs and individuals to act to keep IoT secure.

\subsubsection{Stakeholders in the IoT ecosystem need to work in tandem to secure IoT} \label{lessoncoherenteffort}
Our study shows that an effort to keep the IoT ecosystem secure is profoundly complex.
Although ISPs and individuals are key in securing IoT, it is evident that they cannot do this alone.
Other stakeholders, particularly device-makers and government, also have an important role to play and need to take their shares of responsibilities. 

This is certainly easier said than done, as stakeholders like to blame others for existing issues, at least among the ISPs and individuals examined in this study.
For example, ISPs feel that device makers and individuals do not sufficiently secure IoT devices.
Individuals want ISPs to take some initiatives to secure IoT. They want their traffic to be monitored and have malicious traffic from IoT devices blocked, but the ISPs in Japan are not able to do so unless the government amends the current law.
To tackle these entangling problems, a collaborative effort from relevant parties is a sound approach.
It will not be easy for it requires commitments and enormous effort.
To a certain extent, NOTICE in Japan has set an example of this approach.
The initiative is not perfect; it faces numerous obstacles as shown in this study.
However, the initiative demonstrates that a collaborative effort from key stakeholders in the IoT ecosystem is not impossible.

\subsubsection{ISPs play an important role in keeping IoT secure but it is an uphill battle} \label{lessonisp}
ISPs mostly agree that they are responsible for various activities in keeping IoT secure and feel that they have made contributions, especially in detection, notification, and remediation activities.
This endeavor is not without challenges considering the many internal and external barriers the ISPs are facing.
Asking ISPs to do more, such as establishing various initiatives that individuals want from ISPs will be an uphill battle.

Implementing incentives is key to encouraging ISPs to act.
Internally, most ISPs believe that acquiring technology to identify compromised IoT devices in their network and obtaining executive and staff buy-in are key.
Convincing executive is often not easy, especially if IoT risks (or cybersecurity risks in general) are not well understood at the board level. 
Proactive management of cybersecurity risks at this level is one idea to overcome this difficulty~\cite{ncscboard}.

Externally, ISPs want several incentives from other stakeholders.
Implementing these external incentives are largely beyond the ISPs' control but the ISPs, as a community, can take a proactive approach and influencing it to happen.
For example, starting a conversation with lawmakers about the possibility of changing the law to allow ISPs to monitor and block malicious IoT traffic, and essentially set up a walled garden.
The more challenging issue is that even if ISPs are allowed to offer this service, there is little economic incentive from customers.
Their willingness to pay ISPs for IoT security services is too low for the ISPs to offer the service and increasing individuals' WTP is hard as we discuss in the next lesson.
This is where government subsidies could be tremendously helpful, until ISPs can achieve economies of scale from offering such service.

\subsubsection{Individuals expect more and should be asked to do less to secure IoT}
\label{lessonindividual}
Like ISPs, individuals also wanted several other stakeholders to act to keep IoT secure.
The majority of them in both countries want ISPs to monitor and block malicious traffic to and from IoT devices and want IoT device makers to make IoT devices secure.

However, individuals are not willing to pay much for increased security. The amount that participants in Japan are willing to pay ISPs for IoT security service is not enough to convince ISPs to want to offer the service. 
A large proportion of participants do not want to pay to enhance security service on IoT devices either, suggesting device makers have to absorb the cost of implementing security on IoT devices.
This suggests that individuals do not see these services as having a great value to them.

Therefore, solutions relying less on individuals to keep IoT secure are more promising. 
This does not mean we should completely take individuals out of the equation; they do have their fair share of responsibility---for example, acting to remediate when notified that a device has been compromised. But the conditions must be correct for them to act. They must have motivation, knowledge, and support to do so.

Participants in this study identified being exposed to stories about past attacks on IoT devices as a source of awareness of IoT risks and motivation for keeping their devices secure. It is in the interest of ISPs and governments to help make individuals more aware of these incidents.

\subsection{Government is at the heart of securing IoT}
\label{lessongov}
Although this study focuses on ISPs and individuals, our findings strongly suggest that government is at the core of driving an effort to secure the IoT ecosystem.
A long list of barriers and incentives revolve around the government's role and responsibilities.
Understandably, governments cannot implement all of the identified measures.
We propose that the following incentives be prioritised as they were identified as desirable by a large proportion of ISPs and individuals:

\subsubsection{Allow ISPs to lawfully monitor and block suspicious IoT traffic.}
Our study provided evidence that ISPs and individuals in Japan want the Telecommunications Business Law Article 4 to change to allow ISPs to monitor and block suspicious IoT traffic.

This is in direct conflict with Japan's strong emphasis on data privacy, although privacy-preserving implementations may be possible.
However, the amendment of law to specifically allow NOTICE to scan and identify vulnerable IoT devices nationwide demonstrated that changing the law for the purpose of cybersecurity is indeed possible. Further work could explore support for such laws among individuals when the wider implications (e.g., on privacy) are better understood.

\subsubsection{Provide subsidy to ISPs.}
Government subsidy is another item on the ISPs' wish list. 
A previous study suggested that this form of incentive is effective in encouraging ISPs to clean up malware infection on computers~\cite{clayton2011might}.
Therefore, considering offering the subsidy to keep IoT secure can also be a good idea.
The advantages and disadvantages of doing so need to be analyzed in greater depth.
This goes beyond the scope of this study but evidence from our study is useful to demonstrate the advantage of this incentive.

\subsubsection{Increase the visibility of regulation around IoT.}
\label{tigentenregulaton}
ISPs and individuals want IoT device makers to design and make secure IoT devices.
Tightening the law and regulations around IoT is one way to achieve this (e.g. the UK~\cite{ukcodeofpractice}, Australia~\cite{aucodeofpractice}).
Japan has also made some significant progress.
Effective from 1\st April 2020, all devices must be equipped with sufficient access control, the ability to prompt users to change default passwords, software/firmware updates, and security settings resilient to device rebooting~\cite{iotnewrule}.

Yet, many ISPs and participants in our study still want device makers to make secure devices. Public perception will take time to change if the new laws and regulations are indeed sufficient.
Increasing the visibility of these changes will also give some assurance to individuals about the safety of IoT and boost the confidence in the government's handling of IoT security.

\subsubsection{Give public recognition to ISPs.}
ISPs want public recognition for the contribution they make to keeping IoT secure.
This classic form of economic incentive is not uncommon.
However, it is rather surprising that one in two ISPs in our study want it, considering praising and giving recognition for good work is rare in Japanese culture.
The government needs to consider carefully the appropriate forms of public recognition to be given to ISPs to avoid a possible culture clash.
A more subtle approach is likely to be preferable.

\subsection{Limitations}
\label{limitation}

In the ISPs study, participants may not be fully engaged in the pre-workshop survey or experience survey fatigue, though we did not experience this during our pilot study. The number of ISPs was also relatively small but we consider their mixed demographic and statistically significant results reasonably sufficient to represent the opinion of the ISP community in Japan. 

Our studies relied heavily on self-reporting. We iterated on the surveys and interview guides to encourage accurate reporting by the participants and attempted to minimise leading questions. Interviews were conducted online but took place nearly a year after the pandemic started. Hence, we believe that this method did not considerably affect the quality of the interviews.

The CV method used to evaluate the WTP is also subjected to criticism~\cite{boyle2017contingent}, but it is still being used in economic literature today. For the econometric and statistical analysis, unravelling correlation and causation is problematic and some of the results may reflect influences from underlying variables. Some findings can be hard to generalise, particularly those concerning law and regulation which can vary vastly from one country to another. 

\section{Conclusion}
\label{Conclusion}

We examined the attitudes, barriers, and incentives for ISPs and individuals to keep IoT secure, drawing on evidence from Japan.
Our findings show that although ISPs and individuals have some concern about the security and privacy risks of IoT, they do not prioritise the need to keep IoT secure.
A large number of the incentives and measures we identified are external to ISPs and individuals; 
our study demonstrates that improving the security of IoT will require support from other key stakeholders in the IoT ecosystem.

The complexity of the effort to secure the IoT ecosystem means that there are still many follow-on questions, such as \enquote{What is the best approach for ISPs to increase executive and staff buy-in to invest in keeping IoT secure?} and \enquote{What interventions can increase participants' ability and motivation to remediate their devices?} Like the other stakeholders, the research community has a role to play in this effort.

\bibliographystyle{plain}

\footnotesize
\bibliography{reference}

\begin{thebibliography}{10}

\bibitem{active}
ACTIVE.
\newblock {Advanced Cyber Threats response Initiative}.
\newblock \url{https://www.ict-isac.jp/active/en/active/}, 2015.
\newblock [Online; accessed 22-Feb-2021].

\bibitem{alrawi2019sok}
Omar Alrawi, Chaz Lever, Manos Antonakakis, and Fabian Monrose.
\newblock Sok: Security evaluation of home-based {IoT} deployments.
\newblock In {\em 2019 IEEE symposium on security and privacy ({S\&P})}, pages
  1362--1380. IEEE, 2019.

\bibitem{antonakakis2017understanding}
Manos Antonakakis, Tim April, Michael Bailey, Matt Bernhard, Elie Bursztein,
  Jaime Cochran, Zakir Durumeric, J~Alex Halderman, Luca Invernizzi, Michalis
  Kallitsis, et~al.
\newblock Understanding the mirai botnet.
\newblock In {\em 26th {USENIX} Security Symposium}, pages 1093--1110, 2017.

\bibitem{asghari2015economics}
Hadi Asghari, Michel~JG van Eeten, and Johannes~M Bauer.
\newblock Economics of fighting botnets: Lessons from a decade of mitigation.
\newblock {\em IEEE Security \& Privacy}, 13(5):16--23, 2015.

\bibitem{baskerville1999investigating}
Richard~L Baskerville.
\newblock Investigating information systems with action research.
\newblock {\em Communications of the association for information systems},
  2(1):19, 1999.

\bibitem{cctvhackus}
BBC.
\newblock {Hack of '150,000 cameras' investigated by camera firm}.
\newblock \url{https://www.bbc.com/news/technology-56342525}, 2021.
\newblock [Online; accessed 20-Mar-2021].

\bibitem{blythe2020security}
John~M Blythe, Shane~D Johnson, and Matthew Manning.
\newblock What is security worth to consumers? investigating willingness to pay
  for secure internet of things devices.
\newblock {\em Crime Science}, 9(1):1--9, 2020.

\bibitem{bouwmeester2021thing}
Brennen Bouwmeester, Elsa Rodr{\'\i}guez, Carlos Ga{\~n}{\'a}n, Michel van
  Eeten, and Simon Parkin.
\newblock ``the thing doesn't have a name'': Learning from emergent real-world
  interventions in smart home security.
\newblock In {\em Seventeenth Symposium on Usable Privacy and Security ({SOUPS}
  2021)}, pages 493--512. {USENIX} Association, August 2021.

\bibitem{boyle2017contingent}
Kevin~J Boyle.
\newblock Contingent valuation in practice.
\newblock In {\em A primer on nonmarket valuation}, pages 83--131. Springer,
  2017.

\bibitem{braun2006using}
Virginia Braun and Victoria Clarke.
\newblock Using thematic analysis in psychology.
\newblock {\em Qualitative research in psychology}, 3(2):77--101, 2006.

\bibitem{ccc}
CCC.
\newblock {Cyber Clean Centre}.
\newblock \url{https://www.telecom-isac.jp/ccc/en_index.html}, 2010.
\newblock [Online; accessed 22-Feb-2021].

\bibitem{ncscboard}
National Cyber~Security Centre.
\newblock {10 steps to cyber security - A board level responsibility}.
\newblock
  \url{https://www.ncsc.gov.uk/collection/10-steps-to-cyber-security/introduction-to-cyber-security/board-level-responsibility},
  2021.
\newblock [Online; accessed 15-Mar-2021].

\bibitem{ccetin2019cleaning}
Or{\c{c}}un {\c{C}}etin, Carlos Gan{\'a}n, Lisette Altena, Takahiro Kasama,
  Daisuke Inoue, Kazuki Tamiya, Ying Tie, Katsunari Yoshioka, and Michel van
  Eeten.
\newblock Cleaning up the internet of evil things: Real-world evidence on isp
  and consumer efforts to remove mirai.
\newblock In {\em NDSS}, 2019.

\bibitem{cetin2018let}
Or{\c{c}}un Cetin, Carlos Gan{\'a}n, Lisette Altena, Samaneh Tajalizadehkhoob,
  and Michel van Eeten.
\newblock Let me out! evaluating the effectiveness of quarantining compromised
  users in walled gardens.
\newblock In {\em Fourteenth Symposium on Usable Privacy and Security ({SOUPS}
  2018)}, pages 251--263, 2018.

\bibitem{ccetin2019tell}
Or{\c{c}}un {\c{C}}etin, Carlos Ga{\~n}{\'a}n, Lisette Altena, Samaneh
  Tajalizadehkhoob, and Michel van Eeten.
\newblock Tell me you fixed it: Evaluating vulnerability notifications via
  quarantine networks.
\newblock In {\em 2019 IEEE European Symposium on Security and Privacy
  (EuroS\&P)}, pages 326--339. IEEE, 2019.

\bibitem{cetin2017make}
Orcun Cetin, Carlos Ganan, Maciej Korczynski, and Michel van Eeten.
\newblock Make notifications great again: learning how to notify in the age of
  large-scale vulnerability scanning.
\newblock In {\em Workshop on the Economics of Information Security (WEIS)},
  2017.

\bibitem{clayton2011might}
Richard Clayton.
\newblock Might governments clean-up malware?
\newblock {\em Communication and Strategies}, (81):87--104, 2011.

\bibitem{cohen2004willingness}
Mark~A Cohen, Roland~T Rust, Sara Steen, and Simon~T Tidd.
\newblock Willingness-to-pay for crime control programs.
\newblock {\em Criminology}, 42(1):89--110, 2004.

\bibitem{ukcodeofpractice}
Media Department~for Digital, Culture and Sport.
\newblock {Code of Practice for Consumer IoT Security}.
\newblock
  \url{https://assets.publishing.service.gov.uk/government/uploads/system/uploads/attachment_data/file/773867/Code_of_Practice_for_Consumer_IoT_Security_October_2018.pdf},
  2018.
\newblock [Online; accessed 15-Mar-2021].

\bibitem{dietz2018iot}
Christian Dietz, Raphael~Labaca Castro, Jessica Steinberger, Cezary Wilczak,
  Marcel Antzek, Anna Sperotto, and Aiko Pras.
\newblock Iot-botnet detection and isolation by access routers.
\newblock In {\em 2018 9th International Conference on the Network of the
  Future (NOF)}, pages 88--95. IEEE, 2018.

\bibitem{lazaroreasons}
Steve Dodier-Lazaro, Ingolf Becker, Jens Krinke, and M.~Angela Sasse.
\newblock No good reason to remove features: Expert users value useful apps
  over secure ones.
\newblock In Theo Tryfonas, editor, {\em Human Aspects of Information Security,
  Privacy and Trust}, pages 25--44, Cham, 2017. Springer International
  Publishing.

\bibitem{edwards2011role}
Lilian Edwards.
\newblock The role of internet intermediaries in advancing public policy
  objectives forging partnerships for advancing policy objectives for the
  internet economy, part ii.
\newblock {\em Part II (June 22, 2011)}, 2011.

\bibitem{emami2019exploring}
Pardis Emami-Naeini, Henry Dixon, Yuvraj Agarwal, and Lorrie~Faith Cranor.
\newblock Exploring how privacy and security factor into iot device purchase
  behavior.
\newblock In {\em Proceedings of the 2019 CHI Conference on Human Factors in
  Computing Systems}, pages 1--12, 2019.

\bibitem{aucodeofpractice}
Australian Government.
\newblock {Code of Practice: Securing the Internet of Things for Consumers}.
\newblock
  \url{https://www.homeaffairs.gov.au/reports-and-pubs/files/code-of-practice.pdf},
  2020.
\newblock [Online; accessed 15-Mar-2021].

\bibitem{haney2021company}
Julie Haney, Yasemin Acar, and Susanne Furman.
\newblock ``{It's the Company, the Government, You and I}'': User perceptions
  of responsibility for smart home privacy and security.
\newblock In {\em 30th {USENIX} Security Symposium ({USENIX} Security '21)},
  pages 411--428. {USENIX} Association, August 2021.

\bibitem{kaku1997path}
Ryuzaburo Kaku.
\newblock The path of kyosei.
\newblock {\em Harvard Business Review}, 75:55--64, 1997.

\bibitem{kaperskyiotdyn}
Kate Kochetkova.
\newblock {How to not break the Internet}.
\newblock \url{https://www.kaspersky.com/blog/attack-on-dyn-explained/13325/},
  2016.
\newblock [Online; accessed 21-Mar-2021].

\bibitem{liu2019uncovering}
Haoyu Liu, Tom Spink, and Paul Patras.
\newblock Uncovering security vulnerabilities in the belkin wemo home
  automation ecosystem.
\newblock In {\em 2019 IEEE International Conference on Pervasive Computing and
  Communications Workshops (PerCom Workshops)}, pages 894--899. IEEE, 2019.

\bibitem{japaneu}
Maths Lundin and Sven Eriksson.
\newblock {Internet of Things Market in Japan}.
\newblock
  \url{www.eu-japan.eu/sites/default/files/publications/docs/booklet_web.pdf},
  2015.
\newblock [Online; accessed 25-Feb-2021].

\bibitem{smarthomehack}
Jake Maher.
\newblock {Hacker Takes Over Couple's Smart Home, Plays Vulgar Music And Raises
  Temperature to 90 Degrees}.
\newblock \url{https://www.newsweek.com/google-nest-hack-milwaukee-1460806},
  2019.
\newblock [Online; accessed 21-Mar-2021].

\bibitem{morgan_iterative_2020}
David~L. Morgan and Andreea Nica.
\newblock Iterative thematic inquiry: A new method for analyzing qualitative
  data.
\newblock {\em International Journal of Qualitative Methods}, 19, 2020.

\bibitem{morgner2017insecure}
Philipp Morgner, Stephan Mattejat, Zinaida Benenson, Christian M{\"u}ller, and
  Frederik Armknecht.
\newblock Insecure to the touch: attacking zigbee 3.0 via touchlink
  commissioning.
\newblock In {\em Proceedings of the 10th ACM Conference on Security and
  Privacy in Wireless and Mobile Networks}, pages 230--240, 2017.

\bibitem{notice}
NOTICE.
\newblock {National Operation Towards IoT Clean Environment}.
\newblock \url{https://notice.go.jp/en}, 2019.
\newblock [Online; accessed 26-Feb-2021].

\bibitem{oecdisp}
OECD.
\newblock Proactive policy measures by internet service providers against
  botnets.
\newblock {\em OECD Digital Economy Papers}, 2012.

\bibitem{iotnewrule}
Ministry of~Internal~Affairs and Communications Japan.
\newblock {1985 Post Office Ordinance No. 31, Terminal equipment rules}.
\newblock \url{https://elaws.e-gov.go.jp/document?lawid=360M50001000031}, 2020.
\newblock [Online; accessed 12-Mar-2021].

\bibitem{ozanne2008participatory}
Julie~L Ozanne and Bige Saatcioglu.
\newblock Participatory action research.
\newblock {\em Journal of consumer research}, 35(3):423--439, 2008.

\bibitem{pijpker2016role}
Jeroen Pijpker and Harald Vranken.
\newblock The role of internet service providers in botnet mitigation.
\newblock In {\em 2016 European Intelligence and Security Informatics
  Conference (EISIC)}, pages 24--31. IEEE, 2016.

\bibitem{rowe2013home}
Brent Rowe and Dallas Wood.
\newblock Are home internet users willing to pay isps for improvements in cyber
  security?
\newblock In {\em Economics of information security and privacy III}, pages
  193--212. Springer, 2013.

\bibitem{sasse2005usable}
M~Angela Sasse and Ivan Flechais.
\newblock Usable security: Why do we need it? how do we get it?
\newblock O'Reilly, 2005.

\bibitem{seip1992willingness}
Kalle Seip and Jon Strand.
\newblock Willingness to pay for environmental goods in norway: A contingent
  valuation study with real payment.
\newblock {\em Environmental and Resource Economics}, 2(1):91--106, 1992.

\bibitem{sombatruang_internet_2023}
Nissy Sombatruang, Tristan Caulfield, Ingolf Becker, Akira Fujita, Takahiro
  Kasama, Koji Nakao, and Daisuke Inoue.
\newblock Internet {{Service Providers}}' and {{Individuals}}' {{Attitudes}},
  {{Barriers}}, and {{Incentives}} to {{Secure IoT}}.
\newblock In {\em {{USENIX Security Symposium}}}. {USENIX Association}, 2023.

\bibitem{japaniotstat}
Statistica.
\newblock {Value of the internet of things (IoT) application market in Japan
  from 2018 to 2024}.
\newblock
  \url{www.statista.com/statistics/1026279/japan-application-internet-of-things-market-size/},
  2018.
\newblock [Online; accessed 26-Feb-2021].

\bibitem{iotstat}
Statistica.
\newblock {Forecast end-user spending on IoT solutions worldwide from 2017 to
  2025}.
\newblock \url{
  https://www.statista.com/statistics/976313/global-iot-market-size}, 2021.
\newblock [Online; accessed 20-Mar-2021].

\bibitem{sullivan2013analyzing}
Gail~M Sullivan and Anthony~R Artino~Jr.
\newblock Analyzing and interpreting data from likert-type scales.
\newblock {\em Journal of graduate medical education}, 5(4):541--542, 2013.

\bibitem{cctvhackuk}
The Sun.
\newblock {HACK ATTACK Hackers attack UK school CCTV and stream live footage of
  pupils online}.
\newblock
  \url{https://www.thesun.co.uk/news/5670211/hackers-attack-uk-school-cctv-stream-footage-pupils/},
  2018.
\newblock [Online; accessed 20-Mar-2021].

\bibitem{cctvhackjapan}
The~Japan Time.
\newblock {Hackers disable scores of Canon-made security cameras across Japan}.
\newblock
  \url{https://www.japantimes.co.jp/news/2018/05/07/national/hackers-disable-scores-canon-made-security-cameras-across-japan/},
  2018.
\newblock [Online; accessed 20-Mar-2021].

\bibitem{van2010role}
Michel Van~Eeten, Johannes~M Bauer, Hadi Asghari, Shirin Tabatabaie, and David
  Rand.
\newblock The role of internet service providers in botnet mitigation an
  empirical analysis based on spam data.
\newblock TPRC, 2010.

\bibitem{west2008psychology}
Ryan West.
\newblock The psychology of security.
\newblock {\em Communications of the ACM}, 51(4):34--40, 2008.

\bibitem{zheng_presenting_2022}
Sarah Zheng and Ingolf Becker.
\newblock Presenting suspicious details in {{User-Facing E-mail}} headers does
  not improve phishing detection.
\newblock In {\em Eighteenth Symposium on Usable Privacy and Security
  ({{SOUPS}} 2022)}, pages 253--271. {USENIX Association}, 2022.

\end{thebibliography}

\clearpage

\appendix

\section{Appendices}
\label{Appendix}

\subsection{Background} \label{AppendixBackground}
Advanced Cyber Threats Response Initiative (ACTIVE) ~\cite{active} is a public-private partnership between the Ministry of Communication, ISPs, and security vendors in Japan. The project's goals were preventing malware infection and damage from being infected, and removing malware infection. ACTIVE sends a warning to Internet users to prevent malware infection, removes malware, and encouraged users to take autonomous protection measures against malware infection. The project ran between 2013 and 2018. There were 34 ISPs and 18 companies who participated in the project.

Cyber Clean Centre (CCC)~\cite{ccc} is a public-private partnership between the Ministry of Communication, ISPs, and security vendors in Japan. The project aimed to promote bot removal and prevent the re-infection on Internet users' computers. The CCC captured and analyzed the activities of bot samples and developed bot removal tools. ISPs identified infected PCs and their owners, notified them, and encouraged them to download the disinfect tools which were offered free of charge. The project ran between 2006 and 2011. There were 71 ISPs and 7 security vendors who participated in the CCC. 

\subsection{Participant Demographics}

\begin{table}[H]
\caption{Demographics of individuals}
\centering
\scriptsize
\begin{tabular}[p]{p{4.5cm}p{0.35cm}p{0.35cm}p{0.35cm}p{0.35cm}}
 \toprule
 & \multicolumn{4}{c}{Japan} \\
 Demographic &\multicolumn{2}{c}{\textit{Interviews}} &\multicolumn{2}{c}{\textit{Survey}} \\
 &\textit{n} &\%  &\textit{n} &\% \\
 \midrule
 \textit{Gender} & &   \\
 Male &10 &50  &162 &49  \\
 Female &10 &50  &166 &51  \\
 Total &20 &100  &328 &100  \\
 \midrule
 \textit{Age} & &  & & \\
 18-24 years &5 &25 &78 &24 \\
 25-34 years &5 &25 &79 &24  \\
 35-59 years &5 &25 &80 &24 \\
 $>=60$ years &5 &25 &91 &28  \\
 Total &20 &100 &328 &100 \\
 \midrule
 \textit{IoT device used by participants*} & &  & &  \\
 Smart speaker (e.g. Alexa) &7 &35 &- &-  \\
 Set-top box (e.g. Fire stick) &7 &35 &- &- \\
 Smart Internet connected CCTV &5 &25 &- &-  \\
 Smart game console &11 &55 &- &- \\
 Smart watch &8 &40 &- &-  \\
 Smart TV and recording device &16 &80 &- &-  \\
 Smart digital camera &8 &40 &- &-  \\
 Smart home appliance (e.g. light, fridge) &7 &35 &- &-  \\
 Smart home office (e.g. printer) &1 &5 &- &- \\
 Smartphone &20 &100 &- &-  \\
 Tablet &15 &75 &- &-  \\
 Router &19 &95 &- &-  \\
 \midrule
 Smart speaker (e.g. Alexa) &- &- &107 &33  \\
 Smart TV, incl. set-top box (e.g. Fire stick) &- &- &126 &38  \\
 Smart security (e.g. CCTV, door lock) &- &- &243 &13  \\
 Smart utility meter &- &- &65 &20  \\
 Smart hobby (e.g. games console) &- &- &97 &30  \\
 Smart remote controls (e.g. light, curtain) &- &- &54 &17  \\
 Smart kids devices (e.g. baby monitor) &- &- &16 &5  \\
 Smart health (e.g. fitness tracker, watch) &- &- &74 &23  \\
 Smart kitchen (e.g. fridge) &- &- &51 &16 \\
 Smart gardening (e.g. sprinkler) &- &- &14 &4  \\
 Smart home office (e.g. printer) &- &- &114 &35  \\
 Smartphone or tablet &- &- &255 &78 \\
 Router & & &256 &78  \\
 \bottomrule
 \end{tabular} \vspace{0.3em}
 \label{table:demographicIndividual}
\parbox{\columnwidth}{ \scriptsize
* Based on the pre-screen criteria. 
}
\end{table}

\begin{table}[H]
\caption{Demographics of participating ISPs}
\centering
\scriptsize
\begin{tabular}[p]{p{2.25cm}p{.5cm}p{.5cm}p{.5cm}p{.5cm}p{.5cm}p{.5cm}}
 \toprule
 &\multicolumn{2}{c}{\textit{Questionnaire}} &\multicolumn{2}{c}{\textit{1\st Workshop}} &\multicolumn{2}{c}{\textit{2\nd Workshop}}\\
 &\textit{n} &\%  &\textit{n} &\% &\textit{n} &\% \\
 \midrule
 \multicolumn{7}{l}{{\textit{No. of customers}}} \\
 $<10,000$ &4 &15  &1 &20 &1 &14 \\
 10,000 - 1 million &15 &55  &1 &20 &2 &29\\
 $>1$ million &8 &30  &3 &60 &4 &57\\
 Total  &27 &100  &5 &100 &7 &100\\
 \midrule
 \multicolumn{7}{l}{\textit{Service coverage}} \\
 Regional &9 &33  &2 &40 &3 &43\\
 National &18 &67  &3 &60 &4 &57\\
 Total  &27 &100 &5 &100 &7 &100\\
 \midrule
 \multicolumn{7}{l}{\textit{Participation in government initiatives to promote safe ICT*}}\\
 NOTICE* &27 &100  &5 &100 &7 &100\\
 ACTIVE** &9 &33  &3 &60 &4 &57 \\
 CCC*** &11 &41  &4 &80 &5 &71\\
 \bottomrule
 \end{tabular} \vspace{0.3em}
 \label{table:demographicISP}
\parbox{\columnwidth}{ \scriptsize 
* National Operation Towards IoT Clean Environment (NOTICE), **Advanced Cyber Threats response Initiative (ACTIVE), ***Cyber Clean Centre (CCC)
}
\end{table}

\subsection{Results}

\begin{table}[H]
\caption{\% of ISPs per internal barriers to secure IoT}
\centering
\scriptsize
\begin{tabular}[p]{p{1.7cm}p{0.3cm}p{0.4cm}p{0.3cm}p{0.4cm}p{0.3cm}p{0.4cm}p{0.3cm}p{0.4cm}}
 \toprule
   &\multicolumn{2}{c}{Overall} &\multicolumn{2}{c}{Large} &\multicolumn{2}{c}{Medium} &\multicolumn{2}{c}{Small}\\
No. of ISPs &\multicolumn{2}{c}{(27)} &\multicolumn{2}{c}{(8)} &\multicolumn{2}{c}{(15)} &\multicolumn{2}{c}{(4)} \\
\midrule
Barriers &$n$ &\% &$n$ &\% &$n$ &\% &$n$ &\%\\
  \midrule
Human resource &17 &65 &3 &38 &11 &73 &3 &75\\
Executive buy-in &12 &44 &5 &63 &6 &40 &1 &25\\
Financial &10 &37 &5 &63 &5 &33 &0 &0\\
Technology &10 &37 &0 &0 &7 &47 &3 &75\\
Others$^\dagger$ &6 &22 &4 &50 &1 &7 &1 &25\\
Not sure &1 &4 &0 &0 &1 &7 &0 &0\\
\midrule
$\chi^2(5)$ &\multicolumn{2}{c}{24.13**} &\multicolumn{2}{c}{14.66*} &\multicolumn{2}{c}{21.50**} &\multicolumn{2}{c}{10.50}\\
 \bottomrule
\end{tabular} \vspace{0.3em}
\label{table:ISPbarrier}
\parbox{\columnwidth}{ \scriptsize
{}$^\dagger$ Specified by ISPs as shortage of staff (internal but already part of the HR obstacle), law and regulation constraint (external), and the perception that it should be the responsibilities of device makers and users to keep IoT secured (external), *, ** $p<0.05$, and $p<0.001$, respectively.
}
\end{table}

\begin{table}[H]
\caption{ISPs' ranking of concerns from IoT attacks}
\centering
\scriptsize
\begin{tabular}[p]{p{3.5cm}p{0.7cm}p{0.7cm}p{0.7cm}p{0.7cm}}
 \toprule
  Concern &Overall &Large ISP &Medium ISP &Small ISP\\
  \midrule
Service disruption to customers &\textbf{1\st} &\textbf{1\st} &\textbf{1\st} &2\nd  \\
Social responsibilities$^\dagger$ &2\nd &3\rd &2\nd &\textbf{1\st} \\
Reputation damage &3\rd &2\nd &4\th &4\th \\
Loss of customer and market share &4\th &4\th &3\rd &5\th \\
Incurred financial costs &5\th &5\th &6\th &3\rd \\
Warning from relevant authorities &6\th &6\th &5\th &6\th \\
 \midrule
$n$ &26$^\ddagger$ &8 &14$^\ddagger$ &4 \\
$\chi^2$ (5) &45.01** &7.86 &34.29** &11.00*\\
\bottomrule
\end{tabular} \vspace{0.3em}
\label{table:ISPconcernranking}
\parbox{\columnwidth}{ \scriptsize
{}$^\dagger$ The need to take responsibilities to society as a result of IoT attacks occurred in their network, $^\ddagger$ Excluded 1 medium ISP whose ranking consensus can not be identified, *, ** denotes $p<0.05$ and $p <0.001$.
}
\end{table}

\subsection{ISPs survey} \label{preworkshopsurvey}

\paragraph*{Part I. About the ISP and participants}
\begin{itemize}
    \item ISP-P1Q1: How do you best describe the size of your organisation?
    \begin{itemize}
    \item Less than 10,000 customers
    \item 10,000 or more and less than 1 million
    \item More than 1 million
    \item I don’t want to answer
    \end{itemize}
  \item ISP-P1Q2: What is the extent of service coverage that you are providing?
    \begin{itemize}
    \item Regional (within specific prefectures or areas)
    \item National
    \end{itemize}
  \item ISP-P1Q2: What best describe your function at your organization? 
  \begin{itemize}
    \item Management  -- Technical  -- Legal  -- Sales and marketing 
                \item Others: please specify
    \end{itemize}
  \item ISP-P1Q3: Is your organisation participating in the NOTICE?
  \item ISP-P1Q4: If your organizations participate in the NOTICE. Have you ever been notified and alerted by NOTICE project?
  \item ISP-P1Q5: Did your company participate in the previously implemented ACTIVE project (https://www.ict-isac.jp/active/)?
  \item ISP-P1Q6: Did your company participate in the previously implemented Cyber Clean Center (CCC) project (https://www.telecom-isac.jp/ccc/)?
    \end{itemize}

\paragraph*{Part II. The problems and the root causes} 
\begin{itemize}
   \item ISP-P2Q1: Please rank the following potential impacts of your network hosting infected/vulnerable IoT devices (e.g. devices infected by malicious software or devices having weak security) that your organization is concern.    \begin{itemize}
    \item Reputational damage 
    \item Incurred financial costs
    \item Service disruption to customers
    \item Warning from regulators
    \item Loss of customers and market shares
    \item Social responsibilities as a result of the attacks
    \end{itemize}
  \item Please rate from most unlikely to unlikely, neutral, likely, most likely
    \begin{itemize}
    \item ISP-P2Q2: Attacks by malware that infects IoT devices within the company's network or from the company's network to other companies' networks (including overseas) that significantly hinders the provision of Internet services to users or incurs financial costs for your company. How likely are you to have this happen in the next 12 months? \label{ISP-P2Q2}
    \item ISP-P2Q3: The next 12 months will be an attack by malware that infects IoT devices from other companies' networks (including overseas) that will significantly hinder the provision of Internet services to users or incur financial costs for your company. How likely are you to happen during the period? \label{ISP-P2Q3}
    \end{itemize}
  \item ISP-P2Q4: If the answer are "low" or "very low", what is the reason? \label{ISP-P2Q4}
  \item ISP-P2Q5: To what extent do you feel the ISP  \textit{should be} responsible for the following activities relating to keeping IoT ecosystem secure? (from strongly disagree to somewhat disagree, neutral, somewhat agree, and strongly agree) \label{ISP-P2Q5}
    \begin{itemize}
    \item Prevention: activities initiated by an ISP that can reduce the vulnerability of a user’s device (e.g. closing unnecessary ports on the rental router before sending)
    \item Detection: actions/activities with the aim of identifying threats on the network of an ISP (e.g. malware infected or vulnerable devices)
    \item Notification: actions/activities conducted by an ISP to
inform a customer.
    \item Remediation: actions/activities initiated by an ISP to remove malicious software from a compromised device (e.g. blocking communications from infected devices, offering countermeasures)
    \item Recovery: activities to resolve the effects of botnets and normalize Internet services (e.g. support for recovering damaged functions and personal data on infected devices)
    \end{itemize}
  \item ISP-P2Q6: To what extent is your organization \textit{currently commit} your resource to in the following activities relating to keeping IoT ecosystem secure? (same answer choices as the previous question). \label{ISP-P2Q6}
  \item ISP-P2Q7: Is keeping IoT secure part of your organisation’s priorities? \label{ISP-P2Q7}
    \begin{itemize}
    \item Higher rank (tasks that must be done)
    \item Medium (desirable task)
    \item Lower level (tasks to be done if there is room)
    \item I am not sure
    \end{itemize}
  \item ISP-P2Q8: If your organisation is to implement and run all activities relating to keeping IoT ecosystem secure (from prevention to detection, notification, remediation, and recovery), what are the internal barriers (in your organization) that may prevent you to do so? \label{ISP-P2Q8}
    \begin{itemize}
    \item Financial: my organisation does not currently have budget to do so or it would be very difficult to find the budget.
    \item Human resource: (without hiring new staff) my organisation does not currently have enough staff with expertise to do it.
    \item Technology: (without investing in new technology) my organisation does not currently have sufficient technology to do so.
    \item Internal consensus building issues: There is a lack of material to help executives understand the need for action.
    \item Other: Please specify.
    \item I am not sure.
    \end{itemize}
  \item ISP-P2Q9: To what extent do you feel other stakeholders in IoT ecosystem have taken \textit{sufficient role and responsibilities} in securing IoT ecosystem? (from strongly disagree to somewhat disagree, neutral, somewhat agree, and strongly agree) \label{ISP-P2Q9}
    \begin{itemize}
    \item IoT users – individuals
    \item IoT users – corporate
    \item Other ISPs (apart from your organization)
    \item IoT device makers
    \item Government 
    \item Academia
    \end{itemize}
  \item ISP-P2Q10: To what extent do you feel current relevant laws and regulation has deter or prevent your organisation to do more on keeping IoT ecosystem secure?     \begin{itemize}
    \item Japan Act on Protection of Personal Information (APPI) 
    \item Net neutrality 
    \item Telecommunications Business Law Article 4 (Protection of "secret of communication")
    \item The EU General Data Protection Regulation (GDPR)
    \end{itemize}
\end{itemize}

\paragraph*{Special section related to NOTICE}
 \begin{itemize}
 \item ISP-P2Q11: Upon receiving a list of potentially infected IP address from NOTICE project, please explain the procedure your organization undertake from identifying the owners of the devices to notifying them. If you have not received a list from NOTICE project, please assume that you have and describe the ‘would-be’ procedures. \label{ISP-P2Q11}
 \item ISP-P2Q12: Upon receiving the list of potentially infected IP address from NOTICE project, what are the major challenges your organisation face in identifying individual users (non-corporate) whose IoT devices are identified as infected/vulnerable? \label{ISP-P2Q12}
 \item ISP-P2Q13: What are the major challenges your organisation faced in identifying corporate entities whose IoT devices are identified as infected/vulnerable? \label{ISP-P2Q13}
 \item ISP-P2Q14: What work well (any tools or strategy) to help your organisation to identify devices owners whose IoT devices are identified as infected/vulnerable? \label{ISP-P2Q14}
 \item ISP-P2Q15: What are the major challenges your organisation faced in notifying individual users whose IoT devices are identified as infected/vulnerable? \label{ISP-P2Q15}
 \item ISP-P2Q16: What are the major challenges your organisation faced in notifying corporate entities whose IoT devices are identified as infected/vulnerable? \label{ISP-P2Q16}
 \item ISP-P2Q17: What work well (any tools or strategy) to help your organisation to notify devices owners whose IoT devices are identified as infected/vulnerable? \label{ISP-P2Q17}
 \item ISP-P2Q18: If your organization is not participating in the NOTICE. What are the factors that prevent your company from participating in the NOTICE project? \label{ISP-P2Q18}
 \end{itemize}

\paragraph*{Part III. The incentives}
\begin{itemize}
 \item ISP-P3Q1: Which of the following may help to motivate your organisation to invest or commit more on keeping IoT ecosystem secure? \label{ISP-P3Q1}
    \begin{itemize}
        \item Financial
            \begin{itemize}
            \item Subsidy from government to pay for parts of activities relating to keeping IoT ecosystem secure
            \item Transferring part of investment and/or operational cost to customers
            \item Monetize safer IoT devices as a service to customers (such as assisting consumers to fix their devices)
            \end{itemize}
        \item Technical
            \begin{itemize}
            \item Technology that allow ISPs to detect infected/vulnerable IoT devices without compromising privacy of their customers
            \end{itemize}   
        \item Laws and regulation   
            \begin{itemize}
            \item Changes to data privacy laws and regulations that would allow ISPs to monitor traffic.
            \item Net neutrality is rather optional than mandatory for ISPs to comply.
            \item Changes to data privacy laws and regulations that would allow ISPs to restrict customers’ internet access should their devices are infected.            \end{itemize}
         \item Others   
            \begin{itemize}
            \item Individual IoT users take necessary steps to secure their devices
            \item Corporate IoT users take necessary steps to secure their devices
            \item Other ISPs take equal responsibilities in securing IoT ecosystem
            \item Being recognised as making a significant contribution to promote online safety in Japan
            \item Continuing public-private sector partnership that are not restricted by time frame (e.g. NICT continues to identify defected/vulnerable devices and government pay for call support centre after the end of NOTICE)
            \item IoT device makers design devices that are 'secure by default'.
            \item Data or index measuring how each ISP contribute to keeping IoT ecosystem secure are available.
            \item Research community provide data or insight that can help ISPs to implement and operate activities relating keeping IoT ecosystem secure more effectively and efficiently.
            \item Incentives within the company to motivate personnel to do it.
            \item Other: Please specify.
            \item I am not sure.
           \end{itemize}
    \end{itemize}
 
\item ISP-P3Q2: In NOTICE project, which of the following may help to motivate your organisation to identify and notify customers whose IoT devices are identified as infected or vulnerable? \label{ISP-P3Q2}
    \begin{itemize}
        \item NICT provides more data about the devices.
        \item Technology that can help customers to authenticate the authenticity of notification sent to customers.
        \item Technology that can help ISPs to track the progress of notification sent to customers (as used in the Cyber Clean Centre).
        \item Disclosure of the risk or impact of leaving the device to be alerted to society (to publicly inform in advance how large the problem is and the significance of countermeasures).
        \item Government subsidies to cover costs associated with the NOTICE project.
        \item Support from academia to evaluate and or identify most effective channel to notify customers.
        \item Create and publish data or indicators showing how well various notification methods have helped reduce the number of devices in danger.
        \item Other: Please specify.
     \end{itemize}
\item ISP-P3Q3: Finally, if you have any other opinions or impressions regarding this survey, please write them down. \label{ISP-P3Q3}
\end{itemize}

\subsection{Individuals guided interview questions} \label{interviewscript} 
\paragraph*{Part I. Uses of IoT devices}
\begin{itemize}
  \item What IoT devices do you currently own and use and how long (approximately) have you used them?
  \item Why did you decide to use these IoT devices? What influenced you to? 
  \item Did you buy any of these IoT devices (or play a major part in the decision to buy)?
  \item From your recollection, what factors did you consider when choosing which devices to buy? Why do these factors important to you?
  \item From the list of factors you explained, what are the top 5 factors from most important to least important?
    \begin{itemize}
    \item If security and privacy factor are NOT on top 5, probe the participants why it is not on their list.
    \item If security and privacy factor are on top 3, probe the participants why it is on their list.
    \end{itemize}
  \item Is there any IoT devices you used to own but not anymore? Why?
  \item Is there any IoT devices you don’t currently own but wish to have one? Can you share the reasons you have not acquire them? What prevent you from acquiring it. If you don’t want to share the reasons, it is ok and we can skip the question.
\end{itemize}

\paragraph*{Part II. Security and privacy of IoT}
\begin{itemize}
  \item Are you concern about the security and privacy of any of IoT devices you own? How concern you are on the scale of 1-5 (1= not concern at all, 2 = not that concern, 3 = concern, 4 = somewhat very concern, 5 = very concern).
  \begin{itemize}
  \item If NOT concerned, why are you not concern about it?
  \item If concerned, what concern do you have and why? Is there any particular type of IoT devices and/or security threats that you concern more than another and why?
  \end{itemize}
  \item For the (selected 3) IoT devices you are using or own, please walk me through what you do when you first setup the devices. 
  \begin{itemize}
  \item If setting security and privacy are NOT part of their initial device setup, why did it not occur to you to do it? 
  \item If setting security and privacy functions are part of their initial device setup, why did you decide to do it and what motivate/prompt you to?
  \end{itemize}
  \item When you set up security and privacy settings, how easy or hard was it and why?
  \item Did you have help from anyone (e.g. families, friends) or any sources (e.g. google search, tech blog, device maker’s call centre) when setting it up? How helpful/unhelpful have these sources was?
  \item What would be helpful to make initial security and privacy setting easier for you? 
  \item After the initial setup and installation of the IoT devices, how often do you re-visit or update these privacy and security setting?
  \begin{itemize}
  \item If never, why did it not occur to you to do it?  
  \item If they have:
    \begin{itemize}
    \item What settings did you update? What prompted you to?
    \item Did you experiences any difficulties? How easy or hard was it and why?
    \item Did you have help from anyone (e.g. families, friends) or any sources (e.g. google search, tech blog, device maker’s call centre) when setting it up? How helpful/unhelpful have these sources was?
    \item What could be useful to make the updates of these security and privacy settings easier for you? 
    \end{itemize}
  \end{itemize}
  \item Have any of your IoT devices been compromised/hacked, or have you suspected your devices might be/have been compromised?
  \begin{itemize}
  \item If no, why did you feel that it has never been compromised? 
  \item If yes, 
    \begin{itemize}
    \item How did you find out that it was compromised? (or how did you come to suspected that it was compromised?)
    \item How did you fix (or try to fix) your devices after you found out that it was compromised (or after you suspect that it may have been compromised)?
        \begin{itemize}
        \item If they did NOT try to fix it, why did you decide not to fix it? What prevent/deter you? Was there any problems? What would help to encourage you to fix it?
        \item If they have tried to fix it, How long did it take you to fix it? If it takes quite a while, why? How was your experience with remediation/fixing have been? Did you get help from any sources? What would make it easier for you to fix these devices?
    \end{itemize}
    \end{itemize}
  \end{itemize}
\end{itemize}

\paragraph*{Part III. Incentives}
\begin{itemize}
  \item Have you ever been contacted by ISPs that your devices have been compromised or vulnerable?
  \item If they have been contacted by ISPs,
    \begin{itemize}
    \item What channel did the ISPs reached you? 
    \item How did you responses to the ISPs? What actions did you take?
    \item Any problems you have experiences with the ISPs in the attempt to remediate IoT devices? 
    \item What can (or should) ISPs do more to help you remediate the devices more effectively or efficiently?
    \end{itemize}
  \item If they have never been contacted by ISPs,
    \begin{itemize}
    \item If ISPs do (in the future), what channel of communication would you prefer ISPs to use (that you are likely to pay attentions to)?
    \item What do you expect them to do to help you remediating your IoT devices? 
  \end{itemize}
  \item How would you feel if your ISP offers a service to keep your IoT secure (e.g. monitoring if your devices may be vulnerable or infected, call centres to guide you to fix the devices if found to be compromised or vulnerable)? Would you be interested in taking up such services?
  \item How much (max) are you willing to pay per month to pay for such service? If you are not willing to pay, it is fine and please feel free to say so.
  \item How would you feel if your ISP increase your current Internet price plan to absorb the cost from their internal activities to keep the IoT ecosystem secure? 
  \item How much of the increase (in \% of your current fee or in amount) would you be willing to tolerate? If you will not tolerate and rather switch to a cheaper ISP, it is fine and please feel free to say so.
  \item Is there anything else you think ISPs should do to help you to secure IoT devices?
\end{itemize}

\begin{itemize}
  \item What are your expectation from IoT device maker to keep IoT ecosystem secure?
  \item What are your expectation from relevant ICT policy makers/regulators (e.g. MIC) to keep IoT ecosystem secure?
  \item Is there anything else you think would motivate you to secure IoT devices? 
\end{itemize}  

\subsection{Individual survey}
\label{consumersurvey}
\paragraph*{Part I: Attitudes}
\paragraph{Concern about security and privacy of IoT devices}
\begin{itemize}
 \item IDV-P1Q1 How concerned are you with the potential security and privacy risks of the following IoT devices? (Not concern at all, Not that concern, Neutral/Not sure, Somewhat concern, Very concern)
    \begin{itemize}
        \item Smartphone/tablet
        \item Smart security devices (e.g. door/lock, CCTV, alarms)
        \item Smart gadget for kids (e.g. baby monitors, smart toys, smart pets)
        \item Smart home office (e.g. printer)
        \item Wi-Fi router
        \item Smart utility meter (e.g. water, electricity, gas)
        \item Smart health appliance (e.g. fitness tracker, smartwatch)
        \item Home assistant (e.g.  Google Home, Amazon Alexa/Echo)
        \item Smart hobbies gadgets (e.g. camera, game consoles)
        \item Smart kitchen appliances (e.g. coffee maker, toasters, fridge)
        \item Smart TV
        \item Smart remote controls (e.g. lights, curtain, etc.)
        \item Smart garden appliances (e.g. water sprinkler)
  \end{itemize} 
\end{itemize} 

\paragraph{Nature of concern}
\begin{itemize}
 \item IDV-P1Q2 Please rank your concern (from most to least) about the security and privacy of IoT devices?  
    \begin{itemize}
        \item Financial loss (e.g. theft of credit card info linked to the devices)
        \item Privacy leakage (e.g. eavesdropping on smart TV/cameras, target ads)
        \item Disruption to my day to day activities (e.g. being locked out from my home)
        \item Psychological harm to users (e.g. cyber stalking, blackmailing sensitive info)
        \item Physical harm to users (e.g. overheating smart boiler)
        \item Physical damages to IoT devices or other devices or properties in the house
    \end{itemize} 
\end{itemize}

\paragraph{Perceived likelihood of devices being compromised}
\begin{itemize}
\item IDV-P1Q3 For the IoT devices you are currently using, how do you perceived the overall possibility that any of your devices may have been infected by malware or tampered with? (Most unlikely, Unlikely, I'm not sure/May be, Likely, Most likely)
\begin{itemize}
\item IDV-P1Q3a If the answer is “I’m not sure/I don’t know” or “Most unlikely” or “Unlikely”,why do you feel that way? 
    \begin{itemize}
        \item My devices are working normally
        \item My anti-virus report does not show any suspicious activities
        \item I never get any notification that it has problems
        \item My credit card or bank account (or other financial means) does not show any suspicious transactions
        \item I have not received any strange or suspicious emails
        \item I audited my own devices and see nothing wrong
        \item None of the above: Please explain your rationale
    \end{itemize} 
\item IDV-P1Q3b If the answer is “May be” or “Likely” or “Most Likely”,why do you feel that way? 
    \begin{itemize}
        \item My devices are behaving oddly from time to time
        \item I have known enemies that may want to hack into my devices
        \item I am using old and outdated devices
        \item I think IoT device makers did not make the device secure enough
        \item My credit card or bank account was hacked 
        \item I received strange or suspicious emails
        \item I audited my own devices and found that it was hacked/tempered
        \item None of the above
    \end{itemize} 
\end{itemize}
\end{itemize} 
 
\paragraph{Priority and commitment to keep IoT secure}
\begin{itemize}
\item IDV-P1Q4 In general, what is your priority in keeping your IoT devices secure?
    \begin{itemize}
        \item High (tasks that must be done)
        \item Medium (desirable task)
        \item Lower level (tasks to be done if there is a room)
        \item I'm not sure
    \end{itemize}
\item IDV-P1Q4a If "High", what the is rationale? 
    \begin{itemize} 
        \item I care a lot about security and privacy in general
        \item I think there are many bad people wanting to hack or infect my IoT devices
        \item I think that device makers have NOT done enough care to secure the devices
        \item I think that my ISPs have NOT done enough in monitoring suspicious traffic should my devices become infected
        \item I think government officials and polices are not capable of arresting cybercriminals, so I take matter in my own hands
        \item I have the skills and knowledge to do so
        \item Others: please specify
    \end{itemize}
\item IDV-P1Q4b If "Medium", what the is rationale? 
    \begin{itemize} 
        \item I feel ambivalence about security and privacy in general
        \item I think there are some bad people out there but not too many of them wanting to hack or infect my IoT devices
        \item To a certain extent, I think device makers have sufficiently secure the devices 
        \item To a certain extent, I rely on my ISPs to monitor suspicious traffic and let me know if my IoT devices are infected
        \item To a certain extent, I rely on government official and police to catch cybercriminals if they hack into or infect my devices
        \item Others: please specify
    \end{itemize}
\item IDV-P1Q4c If "Low" or "I'm not sure", what the is rationale? 
    \begin{itemize} 
        \item I generally do not worry about security and privacy of IoT devices
        \item I do not have the knowledge or skills to keep my IoT secure
        \item I do not think anyone would want to hack or infect my IoT devices
        \item I think that device makers have done enough to secure the devices
        \item I rely on my ISPs to monitor suspicious traffic and let me know if my IoT devices are infected
        \item I think government official and police are capable of arresting cybercriminals, so I know I can leave this matter in their hands
        \item Others: please specify
    \end{itemize}    
\end{itemize} 

\paragraph*{Part II: Barriers}
\paragraph{IoT security hygiene and barriers}
\begin{itemize}
\item IDV-P2Q1 When setting up your IoT devices for the first time, which of the following features did you change? 
    \begin{itemize} 
        \item Set up user account and password to restrict access
        \item Change default password of the devices (if it has one)
        \item Update firmware or software
        \item Set firmware or software update to automatic or manual
        \item Setup 2-factor authentication 
        \item Opt-out from third party’s data sharing   
        \item Erase possible old data from the devices, including factory reset   
        \item I do not remember
    \end{itemize}
\end{itemize} 

\paragraph{Barriers to remediating compromised IoT}
\begin{itemize}
\item IDV-P2Q2 Have you tried to fix it? (After IDV-P1Q3b )
    \begin{itemize} 
        \item Yes 
        \item No
    \end{itemize}

\item IDV-P2Q2a If yes, what are the problems you faced when fixing (or trying to fix) your IoT devices? 
    \begin{itemize} 
        \item I did not have any problem at all
        \item It was just too technical for me
        \item Not enough support from device makers
        \item Not enough support from ISPs
        \item Not enough support from my circle of friends and family
    \end{itemize}

\item IDV-P2Q2b If no, why did you not try to fix it?
    \begin{itemize} 
        \item I did not know where to start 
        \item I thought it was easier to replace it with new devices 
        \item It did not worth the stress and/or my time
        \item I did not care much if my IoT devices was compromised 
        \end{itemize}
    \end{itemize} 

\paragraph*{Part III: Incentives}
\paragraph{Internal incentives}
\begin{itemize}
\item IDV-P3Q1 I would be more willing to keep my IoT devices secure if...
    \begin{itemize} 
        \item I have seen news or reports on media about past IoT attacks 
        \item I have seen variety show or celebrities (including YouTuber) talking about the possible cons of using IoT devices
        \item I have seen ads or campaign aiming at raising awareness of the possible downside of using IoT devices
        \item I have direct experience of being victims of IoT attacks (e.g. someone hacked into my devices and steal my personal data).
        \item My close circle of friends and/or family members mentioned about the danger of using IoT devices.
        \item My school or workplace teach me about the possible danger of using IoT devices
        \item I don’t care about security and privacy
    \end{itemize}
\item IDV-P3Q2 Which media channel do you generally pay most attention to? Please rank them in order of preference
    \begin{itemize} 
        \item TV
        \item Radio
        \item Newspaper
        \item Internet channel e.g. YouTube, Twitch
        \item Electronic display at public space 
        \item Paper display at government facilities e.g. city hall
        \item A pamphlet or leaflet
        \item QR codes that I can scan for more info
    \end{itemize}
\end{itemize} 

\paragraph{External incentives}
\begin{itemize}
\item IDV-P3Q3 Various stakeholders play a part in keeping the overall IoT device ecosystem secure. To what extent do you feel other stakeholders in IoT ecosystem in Japan have taken sufficient role and responsibilities in securing IoT ecosystem? (Strongly disagree, Somewhat disagree, Neutral/I'm not sure, Somewhat agree, Strongly agree)
     \begin{itemize} 
        \item IoT users – individual like yourself
        \item IoT users – corporate 
        \item Internet service providers (ISPs)
        \item IoT device makers
        \item Government 
        \item Research institutions 
    \end{itemize}

\item IDV-P3Q4 What do you want ISPs to help you to keep IoT devices secure?
     \begin{itemize} 
        \item (For rental Wi-Fi router from ISPs) Do proper security setting on the Wi-Fi routers before shipping/giving them to customers
        \item (For rental Wi-Fi router from ISPs) Notify users of critical updates
        \item Monitor suspicious traffic from IoT devices (e.g. traffic from IoT malware) and notify customers before blocking
        \item Monitor suspicious traffic from IoT devices (e.g. traffic from IoT malware) and automatically block such traffics 
        \item Offer a service to fix infected IoT devices
        \item Offer a service to visit customer at home and fix infected IoT devices
        \item Have a call centre that customers can easily talk to human staff
        \item Educate users about the potential danger of using IoT devices
        \item Be transparent about how ISP are protecting customer communication and personal information in relation to IoT ecosystem
    \end{itemize}

\item IDV-P3Q5 If ISPs are to contact you about your IoT being vulnerable or compromised, which channel would you prefer? Please rank them in order of preference.
     \begin{itemize} 
        \item Phone call
        \item SMS
        \item Postcard
        \item Letters
        \item Carrier emails
        \item Personal emails
        \item Home visit
     \end{itemize} 

\item IDV-P3Q6 How much are you willing to pay ISPs for the following services. Please put the max amount/month I will pay (put \textyen0 if you don’t want to pay).
    \begin{itemize} 
        \item A service to monitor suspicious traffic from IoT devices and notify customers
        \item A service to remotely guide users how to fix infected IoT devices
        \item A service to physically (home visit) fix infected IoT devices
        \item A complete package which include monitor suspicious traffic, notify customers, and help the customers to fix the devices
    \end{itemize} 

\item IDV-P3Q7 What do you want IoT device makers to do in order to secure IoT? 
     \begin{itemize} 
        \item Make products secure by default
        \item Make security and privacy settings on IoT devices easy and standardised
        \item Be transparent about their products; informing customers both the pros AND cons
        \item Provide regular software/firmware update
        \item Provide well documented and easy to understand user manual
        \item Have a physical store front (or offer service at re-seller such as Denki store) to help me setup or change privacy and security settings
        \item Have a call centre that I can reach out if I have problems related to devices security
        \item Offer service to fix IoT devices if it is infected with malware 
        \item Others: Please specify
    \end{itemize}

\item IDV-P3Q8 What do you want Japanese government to do to keep IoT ecosystem in Japan secure?
    \begin{itemize} 
        \item Stringent IoT regulation 
        \item Raise public awareness the potential danger of using IoT devices
        \item Educate the public about how to use the IoT device safely
        \item Made necessary changes to the law (Article 4 Secrecy Act) to allow ISPs to monitor suspicious Internet traffics and block such traffics accordingly
        \item Arrest more cybercriminal to set example and deter bad people from committing cybercrime
        \item Promote/urge research community to communicate/educate the public about the risks of using IoT device
    \end{itemize} 
\end{itemize} 

\clearpage

\label{sec:codebooks}

\begin{table*}[!t]
    \subsection{Codebooks}
    \caption{Code book and coding frequency for the free text analysis of the ISP Survey. Participants were asked about challenges they face when notifying individuals and corporate entities. Some codes apply to both individuals and corporate customers.}
    \label{tab:ISPCoding}
    \footnotesize
    \centering
    \setlength\tabcolsep{0.3em} %
    \begin{tabular}{p{4cm}p{10.8cm}cc}
    \toprule
        Category & Challenge & Individual & Corporate \\ \midrule
        Awareness/understanding of IoTs risks and/or NOTICE & Users do not understand what NOTICE is about or the security problem with IoTs device entailed (and it is difficult for them to understand). & 5 & 3 \\ \midrule
        Channel of communication & Do not have up-to-date info about user's contact or info about communication channel (e.g. email) that users use actively & 4 & 1 \\ 
         & Unable to track if the users read the notification either by email or letters & 2 & 2 \\ 
         & Users avoid contacts by ISPs e.g. do not pick up the phone or not answer when visit their house or difficult to reach by phone during business hours & 2 &   \\ 
         & Users do not pay attention to the notification or not responses to notification & 2 & 1 \\ 
         & Users do not trust the notification (and some raised complaints) & 2 & 1 \\ \midrule
        Delay in remediation & Customers do not have maintenance department &   & 1 \\ 
         & Taking long time for customers to action (i.e. to remediate it) or when ISPs do not maintain the devices (i.e. only provide network service) &   & 2 \\ \midrule
        Financial & Associated costs to ISP if notify by letters or visit customers house & 1 &   \\ 
         & Users are reluctant to remediate when the cost is involved & 1 & 1 \\ \midrule
        Identifying responsible parties & A person using IoTs devices are different from the person who is in contract with ISPs; hence, communicating the problem with IOTs devices become difficult & 3 &   \\ 
         & Corporations outsource the maintenance of IT devices to other company (hence, the notification doesn't reach the outsourced company) &   & 2 \\ 
         & In case of apartment complex, users do not always have contract with ISPs. Therefore,  it is not possible for ISPs to identify the real or the exact users and has to rely on property owner or property management company. &   & 1 \\ 
         & Not sure which department/or an in-charge within the user company are dealing or should be dealing with the identified vulnerable/infected IoTs &   & 3 \\ 
         & The actual users of the devices and the person setting up the device may not always be the same person so remediation action e.g. change the config is not possible &   & 1 \\ \midrule
        Lack of skills & Users do not have enough knowledge or skills to fix the devices (e.g. some users also outsourced the maintenance of the devices to third party) & 1 &   \\ \midrule
        Others & Difficulty in accessing the devices due to geographic distances between the users and the devices. & 1 &   \\ \midrule
        System complexity (corporate) & No way to deal with EOS (End of sales) and EOL (End of life) devices, or legacy system &   & 1 \\ 
         & There is a complex relationship among the systems, and it is not easy to fix them (verification is needed, etc.) &   & 1 \\ \midrule
        User ignoring problems and not promptly fix it. & Users ignore the problem and not remediate their devices e.g. if the effect (from having infected/vulnerable IoTs) seems trivial to them & 3 & 2 \\ \bottomrule
    \end{tabular}
\end{table*}

\clearpage
{\onecolumn
\scriptsize
\setlength\tabcolsep{0.18em}
\newlength\umainspace %
\setlength\umainspace{1.2cm} %
\newlength\usubspace %
\setlength\usubspace{2cm} %
\LTcapwidth=\textwidth
 \begin{longtable}[c]{p{\umainspace}@{\hspace{5pt}}p{\usubspace}@{\hspace{5pt}}p{5.5cm}@{\hspace{5pt}}|@{\hspace{5pt}}cccccccccccccccccccc}
 \caption{Code book and coding matrix of the user interviews. A dot (•) indicates that the code was discussed in the interview. For codes regarding incentives, a dot indicates a positive preference of the user, a cross (x) indicates that the user did not prefer it.\label{tab:userInterviewCoding}}\\
 \toprule
Main part & Sub part & Main theme & P1 & P2 & P3 & P4 & P5 & P6 & P7 & P8 & P9 & P10 & P11 & P12 & P13 & P14 & P15 & P16 & P17 & P18 & P19 & P20 \\ 
 \midrule
 \endfirsthead

 \toprule
 \multicolumn{23}{c}{Continuation of Table~\ref{tab:userInterviewCoding}}\\
 \midrule
Main part & Sub part & Main theme & P1 & P2 & P3 & P4 & P5 & P6 & P7 & P8 & P9 & P10 & P11 & P12 & P13 & P14 & P15 & P16 & P17 & P18 & P19 & P20 \\ 
 \midrule
 \endhead

 \bottomrule
 \endfoot

 \bottomrule
 \endlastfoot
 
        \multirow{2}{\umainspace}{Reason for uptake} & \multirow{1}{\usubspace}{Reason for uptake} & Curiosity & • & • &   & • & • & • &   &   & • &   &   &   &   &   &   &   &   &   &   &   \\ 
          &   & Utility & • & • & • &   & • &   & • & • & • & • & • & • & • & • & • & • & • & • & • & • \\ 
          &   & Others (e.g. freebies, on-sale) & • & • &   & • &   &   &   &   & • & • &   & • &   &   &   &   &   &   & • &   \\ \cmidrule{2-23}
          & \multirow{2}{\usubspace}{Factors influencing buying decisions} & Prices &   & • & • & • &   & • & • &   &   &   &   & • & • &   & • &   & • &   & • & • \\ 
          &   & Features and functionality & • & • & • &   &   &   & • & • & • &   & • &   &   &   & • &   & • &   &   & • \\ 
          &   & Product review rating &   &   & • &   &   & • & • & • &   &   &   &   &   & • &   & • & • &   &   &   \\ 
          &   & Friends and family recommendation &   &   &   &   &   & • &   &   &   &   &   &   &   &   &   &   &   &   &   &   \\ 
          &   & Security and privacy features &   &   &   &   & • & • &   & • &   &   &   &   & • & • &   & • &   & • &   &   \\ 
          &   & Brand reputation & • &   & • & • & • &   &   & • &   &   &   &   & • &   &   & • &   & • &   &   \\ 
          &   & Country of origin &   &   & • & • &   &   &   &   &   &   &   &   &   &   &   & • &   &   &   &   \\ 
          &   & Perceived usability (look like easy to use) &   &   &   & • &   &   &   &   & • &   &   &   &   & • &   &   &   & • & • &   \\ 
          &   & Look and feel (appearance) &   &   &   &   &   & • &   & • &   &   &   &   &   & • &   &   &   &   &   &   \\ 
          &   & Compatibility with existing products &   &   &   &   &   &   &   &   &   &   &   &   &   &   & • &   &   &   &   &   \\ 
          &   & Recommendation by shop assistant &   &   &   &   &   &   &   &   &   &   &   &   &   &   &   &   &   &   &   & • \\ \midrule
        \multirow{4}{\umainspace}{Attitude about security and privacy} & Believe that Japanese products are more secure & Believe that Japanese products are more secure &   &   & • & • &   &   &   &   &   &   & • &   &   &   &   & • &   &   &   &   \\ \cmidrule{2-23}
          & \multirow{2}{\usubspace}{Concern about security} & Generally not concern &   & • &   &   & • &   &   &   &   & • &   & • &   &   & • & • & • & • & • &   \\ 
          &   & Have some concern & • &   & • & • &   & • & • & • & • &   & • & • & • & • & • & • &   & • & • & • \\ 
          &   & Depends on the type of devices & • &   &   &   &   &   &   &   &   &   &   & • &   &   & • & • &   & • & • &   \\ \cmidrule{2-23}
          & \multirow{3}{\usubspace}{Reasons for NOT CONCERN about security and privacy} & Unaware of security and privacy risks & • & • &   &   &   & • &   &   &   &   &   &   &   &   &   &   &   &   &   &   \\ 
          &   & False sense of security and optimism biases &   &   &   &   &   &   & • &   & • & • & • &   & • &   &   &   & • & • &   &   \\ 
          &   & Simply don't care or think about it or lack sense of privacy (nothing to hide attitude) &   & • & • &   &   &   &   &   &   & • &   &   &   &   &   &   & • & • & • &   \\ 
          &   & Reliance on device maker &   &   &   &   & • &   &   & • &   &   &   &   &   &   & • &   &   &   &   &   \\ \cmidrule{2-23}
          & \multirow{3}{\usubspace}{Reasons for CONCERN about security and privacy} & Financial loss &   &   & • &   &   & • &   & • & • &   & • &   &   &   & • &   &   &   &   & • \\ 
          &   & Privacy leakage  & • &   & • &   &   &   & • &   & • &   & • & • & • & • & • & • &   &   &   &   \\ 
          &   & Harm to users - physical and psychological & • &   &   &   &   & • & • &   &   &   &   & • & • &   & • &   &   &   &   &   \\ \cmidrule{2-23}
          & \multirow{2}{\usubspace}{Awareness of cybersecurity threats} & Have seen news about cybersecurity breaches &   & • &   & • & • &   &   & • &   & • & • &   & • &   &   & • &   &   & • &   \\ 
          &   & Feeling that there are hacker out there &   &   &   &   &   &   &   &   &   &   &   &   &   & • &   &   &   &   &   &   \\ 
          &   & Feeling that data breaches are now norm &   &   & • &   &   &   &   &   &   &   &   &   &   &   &   &   &   &   &   &   \\ \midrule
        \multirow{5}{\umainspace}{Perceived possibility of IoTs devices being compro\-mised.} & \multirow{2}{\usubspace}{Likelihood of IoTs devices being compromised} & I’m not sure/I don't know &   &   &   &   &   &   &   &   &   &   &   &   &   &   &   &   & • &   &   &   \\ 
          &   & Most unlikely, unlikely & • & • & • & • & • & • & • & • & • & • & • & • & • & • & • & • &   & • & • & • \\ &&&& \\ \cmidrule{2-23}
          & \multirow{4}{\usubspace}{Rational for perceiving IoTs devices unlikely to have been infected.} & Devices working normally & • & • & • & • &   & • &   &   &   &   & • &   &   &   & • &   &   &   & • &   \\ 
          &   & No suspicious activities detected by Anti-virus software or IoTs devices  &   &   &   &   & • & • & • &   & • &   & • &   & • & • & • &   &   & • &   &   \\ 
          &   & No suspicious activities  found on my emails or financial statements &   &   &   &   &   &   &   & • &   &   &   &   &   & • &   &   &   &   & • &   \\ 
          &   & Performed owns audit and found nothing wrong (e.g. no suspicious traffic) &   &   &   &   &   &   & • &   &   &   &   & • &   &   &   & • &   &   &   & • \\ 
          &   & Others (trust in device maker or trust themselves in own security hygiene) &   &   &   &   &   &   &   &   &   & • &   &   &   &   &   &   &   &   & • &   \\  \midrule
        \multirow{3}{\umainspace}{Initial setup of IoTs devices} &  \multirow{2}{\usubspace}{Initial setup of IoTs devices} & Access control &   &   & • &   &   & • &   & • &   &   &   & • & • &   & • &   &   &   & • &   \\ 
          &   & Software update & • &   &   &   &   &   &   &   & • &   &   &   &   &   &   &   &   &   &   &   \\ 
          &   & Privacy setting &   &   & • &   &   &   &   & • &   &   &   &   &   &   & • &   &   &   &   &   \\ 
          &   & Others &   &   &   &   &   &   & • &   &   &   &   &   &   &   &   & • &   &   &   &   \\ 
          &   & I don't remember & • &   &   &   &   &   &   &   &   &   &   &   & • &   &   &   &   &   &   & • \\  \cmidrule{2-23}
          & \multirow{4}{\usubspace}{Experience with setting security and privacy on IoTs devices} & Experience difficulties &   &   & • & • &   &   &   &   &   &   &   & • &   &   & • & • &   &   & • & • \\ 
          &   & Easy (or not that hard) &   &   & • &   & • & • & • & • & • & • &   & • &   &   & • &   &   &   &   &   \\ &&&& \\ &&&& \\  \cmidrule{2-23}
          & \multirow{4}{\usubspace}{Reasons for NOT setting security and privacy during initial setup} & Not mandatory steps (device still working without) &   & • &   &   &   &   &   &   &   &   &   &   &   &   &   &   &   &   &   &   \\ 
          &   & No need to segregate access in the household & • &   &   &   &   &   &   &   &   &   &   &   &   &   &   &   &   &   &   &   \\ 
          &   & Lack knowledge about security and privacy settings &   &   &   &   & • &   &   &   &   &   & • &   &   &   &   &   &   & • &   &   \\ 
          &   & Trust in device maker &   &   &   &   & • &   &   &   &   &   &   &   &   &   &   &   &   &   &   &   \\ 
          &   & Didn't pay attention &   &   &   &   &   &   &   &   &   & • &   &   &   & • &   &   &   &   &   &   \\ \cmidrule{2-23}
          & \multirow{2}{\usubspace}{Automatic vs manual software update} & Automatic update &   & • & • & • &   &   &   & • &   & • &   &   &   & • & • &   &   &   &   &   \\ 
          &   & Manual update &   &   & • & • & • &   &   &   &   &   &   &   &   &   &   & • &   & • &   &  \\ &&&& \\ \cmidrule{2-23}
          & \multirow{2}{\usubspace}{Update software on router} & Never &   & • &   &   &   &   &   &   &   &   &   &   & • & • &   &   &   &   & • &   \\ 
          &   & Yes &   &   &   &   &   &   & • &   &   &   &   & • &   &   &   & • &   &   &   &   \\ \cmidrule{2-23}
          & \multirow{3}{\usubspace}{Changes to security and privacy settings after initial setup} & Access control &   &   &   &   &   &   &   & • &   &   &   &   &   &   &   &   &   &   &   &   \\ 
          &   & Software update &   &   & • & • & • &   & • &   & • & • &   & • & • &   &   & • &   & • &   &   \\ 
          &   & Review logs &   &   &   &   &   &   & • &   &   &   &   &   &   &   &   & • &   &   &   &   \\ 
          &   & I don't remember & • &   &   &   &   &   &   &   &   &   &   &   &   &   &   &   &   &   &   &   \\ \cmidrule{2-23}
          & \multirow{4}{\usubspace}{User support that will make security and privacy setting easier} & User manual &   & • &   &   &   & • & • &   &   &   &   & • &   &   &   & • &   &   &   & • \\ 
          &   & Video demo &   &   &   &   &   &   &   &   &   &   &   &   &   & • &   &   &   &   &   &   \\ 
          &   & Call centre  &   &   & • &   &   &   &   & • &   &   &   &   &   &   &   &   &   &   &   & • \\ 
          &   & Chat function  &   &   & • &   &   &   &   &   &   &   &   &   &   &   &   &   &   &   &   &   \\ 
          &   & Make software update auto or manual by default &   &   &   &   &   &   &   &   & • &   &   & • & • &   & • &   &   &   &   &   \\ 
          &   & Make changing default password of router easier &   &   &   &   &   &   &   &   &   &   &   &   &   &   &   &   &   &   & • &   \\ \cmidrule{2-23}
          & \multirow{3}{\usubspace}{Prior experience of security incidents or scam} & Have prior experience with computer security incident (virus) or scam &   &   &   & • &   & • &   &   & • & • &   &   &   &   &   & • &   &   &   &  \\ &&&& \\ \cmidrule{2-23}
          & \multirow{2}{\usubspace}{Other security hygiene in general} & Did check for sign of spam/malicious emails &   &   &   &   &   &   &   & • & • & • & • &   &   &   &   &   &   &   &   &   \\ 
          &   & Always update software/use latest OS &   &   &   &   &   &   &   &   & • & • &   &   &   &   &   & • &   &   &   &   \\ 
          &   & Safe Wi-Fi practice (e.g. Use a very long Wi-Fi password, don't connect to WEP) &   &   &   &   &   &   &   &   &   &   &   & • &   &   &   & • &   &   &   &   \\ 
          &   & Do not visit dodgy website &   &   &   &   &   &   & • &   &   &   &   & • &   &   &   & • &   &   &   &   \\ 
          &   & Background and interest in security &   &   &   &   &   &   & • &   &   &   &   & • &   &   &   & • &   &   &   & • \\ \midrule
        \multirow{5}{\umainspace}{Common incentives between consumers and ISPs} & \multirow{3}{\usubspace}{Previous contact with ISPs about infected IoTs device} & No & • & • & • & • & • & • & • & • & • & • & • & • & • & • & • & • & • & • & • & • \\ 
          &   & Yes &   &   &   &   &   &   &   &   &   &   &   &   &   &   &   &   &   &   &   & \\ &&&&  \\ \cmidrule{2-23}
          & \multirow{3}{\usubspace}{Preference for future channel of contacts from ISPs} & Phone call & • & x & x & • & • & • & • & x & • &   &   & • & • & x & • & • & x & • &   & • \\ 
          &   & SMS &   & x &   &   &   & • & x & • & • &   &   &   &   &   &   & x &   &   &   &   \\ 
          &   & Postcard &   & • &   & x &   & • & • & x &   &   &   &   &   & • &   &   &   &   &   & • \\ 
          &   & Letters & x & • &   &   &   &   &   &   &   &   &   &   &   & • & • &   &   &   & • &   \\ 
          &   & Carrier emails &   & x &   &   &   & x & x & • & • &   &   & x & x &   &   &   & • &   & x &   \\ 
          &   & Personal emails &   & x &   &   &   &   & x & • &   &   &   & • & • &   &   &   &   &   & • &   \\ 
          &   & Unspecified emails & • & x & • &   &   &   & x &   &   &   & • &   &   &   & • & • &   & • &   & • \\ 
          &   & Home visit &   &   &   &   &   &   &   &   &   &   &   &   & • &   &   &   &   &   &   &   \\ 
          &   & ISP website &   &   &   &   &   &   &   &   &   &   &   &   &   &   &   & • &   &   &   &   \\ \cmidrule{2-23}
          & \multirow{2}{\usubspace}{Willingness to pay ISPs for security} & Will not pay &   &   &   & • &   &   &   &   &   &   &   &   &   &   &   &   &   &   &   &   \\ 
          &   & Pay < = 500yen & • & • &   &   &   &   & • & • & • & • &   & • &   & • & • &   & • &   &   & • \\ 
          &   & Pay >500yen &   &   & • &   & • & • &   &   &   &   & • &   & • &   &   & • &   & • &   &   \\ \cmidrule{2-23}
          & \multirow{5}{\usubspace}{Preference for security services from ISPs whether OPTIONAL or part of current plan} & Optional & • & • &   &   &   & • & • & • &   & • & • &   &   &   &   &   & • &   &   &   \\ 
          &   & Part of current plan &   &   & • &   &   &   &   &   &   &   &   &   &   & • &   &   &   &   &   & • \\ 
          &   & I don’t mind either way &   &   &   &   &   &   &   &   & • &   &   & • & • &   &   & • &   & • & • &  \\ &&&& \\ &&&&  \\ \cmidrule{2-23}
          & Incentives from ISPs & Secure Wi-Fi router &   &   &   &   &   &   &   &   &   &   &   &   &   &   &   & • &   &   & • &   \\ 
          &   & On-going security monitoring &   &   &   &   &   & • & • & • & • &   & • & • & • &   & • & • & • & • & • & • \\ 
          &   & Customer support & • & • &   &   & • & • & • &   & • & • &   &   & • & • & • &   & • & • & • & • \\ 
          &   & Be transparent about what ISPs are doing & • &   &   &   &   &   &   &   &   &   &   &   & • &   &   & • &   & • &   &   \\ \midrule
        \multirow{3}{\umainspace}{Incentives from device makers} & \multirow{2}{\usubspace}{Incentives from device makers} & Secure the product &   & • & • & • & • & • & • & • & • & • & • & • & • & • & • &   &   & • & • & • \\ 
          &   & Customer support &   &   & • &   &   &   & • & • &   &   &   &   &   &   &   & • &   &   &   & • \\ 
          &   & Be transparent about the product and inform users & • &   &   &   &   &   &   & • &   & • &   & • & • & • &   &   & • &   &   &   \\ \midrule
        \multirow{3}{\umainspace}{Incentives from government} & \multirow{2}{\usubspace}{Incentives from government} & Stringent law and regulation on IoTs & • &   &   &   & • &   &   &   & • & • & • &   & • &   & • &   & • & • & • &  \\ 
          &   & Raise public awareness of IoTs risks & • & • & • & • &   & • & • & • &   &   &   &   & • & • & • & • &   &   &   & • \\ &&&& \\ \midrule
        \multirow{2}{\umainspace}{Internal incentives} & Internal incentives & Increase understanding of risks of using IoTs devices & • & • & • & • &   & • &   & • & • & • & • & • & • &   &   & • & • &   & • & • \\ 
          &   & Direct experience of falling victims &   &   &   &   & • &   &   &   &   &   &   &   &   &   &   &   &   &   &   &   \\ 
          &   & Platform to check the status of data breach or suspect traffic &   &   &   &   &   &   &   &   &   &   &   &   &   &   & • &   &   & • &   &   \\ \cmidrule{2-23}
          & \multirow{4}{\usubspace}{Preference for communication channel for security campaign} & TV &   & • &   & • &   &   &   & • & • & x & • &   &   &   &   &   &   &   &   & • \\ 
          &   & Radio &   &   &   &   &   &   &   & • &   &   &   &   &   &   &   &   &   &   &   &   \\ 
          &   & Newspaper &   & x & x & • &   &   & • & • & • &   &   &   &   &   &   &   &   &   &   & • \\ 
          &   & Internet channel e.g. YouTube, Twitch & • &   & • &   &   & • & • &   &   & • &   &   &   & • &   &   &   &   &   &   \\ 
          &   & Electronic display at public space e.g. shopping mall, train station, etc. &   &   & x &   &   &   &   &   &   &   &   &   &   & • &   &   &   &   &   &   \\ 
          &   & Paper display at government facilities e.g. city hall &   &   &   & x &   &   & • &   &   &   &   &   &   &   &   &   &   &   &   &   \\ 
          &   & QR codes that I can scan for more info &   &   &   &   &   &   &   &   &   & • &   &   &   &   &   &   &   &   &   &   \\ 
          &   & Trailer in the cinema &   &   &   &   &   &   &   &   &   &   &   &   &   & • &   &   &   &   &   &   \\ 
          &   & Letter &   &   &   &   &   &   &   &   &   &   &   &   &   &   &   &   &   &   &   & • \\ 
 \end{longtable}
 \twocolumn}

\end{document}